\documentclass[fleqn,usenatbib]{mnras}

\usepackage{newtxtext,newtxmath}

\usepackage[T1]{fontenc}

\DeclareRobustCommand{\VAN}[3]{#2}
\let\VANthebibliography\thebibliography
\def\thebibliography{\DeclareRobustCommand{\VAN}[3]{##3}\VANthebibliography}

\usepackage{graphicx}	
\usepackage{amsmath}	
\usepackage{color}
\usepackage{rotating}
\usepackage{pdflscape}
\usepackage[export]{adjustbox}
\usepackage{hyperref}
\usepackage{CJKutf8}
\usepackage[normalem]{ulem}

\defcitealias{yue_novel_2024}{Y24}
\title[Decomposing quasar radio emission II]{A novel Bayesian approach for decomposing the radio emission of quasars: II. Link between quasar radio emission and black hole mass}

\author[B.-H. Yue et al.]{
B.-H. Yue~\begin{CJK*}{UTF8}{gbsn}(岳博涵)\end{CJK*},$^{1,2}$\thanks{E-mail: bohan.yue@ed.ac.uk (BY)}
K. J. Duncan,$^{1}$
P. N. Best,$^{1}$
M. I. Arnaudova,$^{3}$
L. K. Morabito,$^{4,5}$
J. W. Petley,$^{2,4}$
\newauthor
H. J. A. R\"ottgering,$^{2}$
S. Shenoy,$^{3}$
D. J. B. Smith$^{3}$
\\
$^{1}$Institute for Astronomy, University of Edinburgh, Edinburgh EH9 3HJ, UK\\
$^{2}$Leiden Observatory, Leiden University, PO Box 9513, NL-2300 RA Leiden, The Netherlands\\
$^{3}$Centre for Astrophysics Research, University of Hertfordshire, College Lane, Hatfield AL10 9AB, UK\\
$^{4}$Centre for Extragalactic Astronomy, Department of Physics, Durham University, Durham DH1 3LE, UK\\
$^{5}$Institute for Computational Cosmology, Department of Physics, University of Durham, South Road, Durham DH1 3LE, UK\\
}

\date{Accepted XXX. Received YYY; in original form ZZZ}

\pubyear{2025}

\begin{document}
\label{firstpage}
\pagerange{\pageref{firstpage}--\pageref{lastpage}}
\maketitle

\begin{abstract}
Whether the mass of supermassive black hole ($M_\mathrm{BH}$) is directly linked to the quasar radio luminosity remains a long-debated issue, and understanding the role of $M_\mathrm{BH}$ in the evolution of quasars is pivotal to unveiling the mechanism of AGN feedback. In this work, based on a two-component Bayesian model, we examine how $M_\mathrm{BH}$ affects the radio emission from quasars, separating the contributions from host galaxy star formation (SF) and AGN activity. By modelling the radio flux density distribution of Sloan Digital Sky Survey (SDSS) quasars from the LOFAR Two-metre Sky Survey Data Release 2, we find no correlation between $M_\mathrm{BH}$ and SF rate (SFR) at any mass for quasars at a given redshift and bolometric luminosity. The same holds for AGN activity across most $M_\mathrm{BH}$ values; however, quasars with the top 20\% most massive SMBHs are 2 to 3 times more likely to host strong radio jets than those with lower-mass SMBHs at similar redshift and luminosity. We suggest defining radio quasar populations by their AGN and SF contributions instead of radio loudness; our new definition unifies previously divergent observational results on the role of $M_\mathrm{BH}$ in quasar radio emissions. We further demonstrate that this radio enhancement in quasars with the 20\% most massive SMBHs affects only the $\sim5\%$ most radio bright quasars at a given redshift and bolometric luminosity. We discuss possible physical origins of this radio excess in the most massive and radio-bright quasar population, which remains an interest for future study.
\end{abstract}

\begin{keywords}
quasars: general – quasars: supermassive black holes – galaxies: active – radio continuum: galaxies – black hole physics.
\end{keywords}



\section{Introduction}
\label{sec:intro}

Quasars (QSOs) emit powerful emission across the entire electromagnetic spectrum, ranging from X-rays to radio bands. Yet different components of the quasar dominate at different wavelengths. Historically, since the compilation of some of the earliest optically selected quasar catalogues, it was observed that only a small percentage (about 5 to 10\%) of the optically-selected quasars are also luminous in radio bands when compared to the AGN emission in optical bands \citep[see e.g.][]{kellermann_vla_1989}, which are then referred to as radio-loud (RL). 
The remaining optical quasars that are relatively faint in the radio bands are classified as radio-quiet (RQ). Early studies considered the RL and RQ quasars as two distinct populations with different physical origins \citep[e.g.][]{ivezic_optical_2002,white_signals_2007}. However, more recent studies have disputed this point of view, arguing that radio emissions in RL and RQ quasars share a similar origin, and RL quasars are simply the long tail toward the radio-bright end in a continuous quasar radio power distribution \citep[e.g.][]{cirasuolo_radio-loudradio-quiet_2003,balokovic_disclosing_2012,macfarlane_radio_2021}. \par

Efforts to tackle the questions behind a possible quasar radio bimodality mainly focused on investigating the possible differences in the physical properties of the central engines to quasar radio emission: supermassive black holes (SMBHs). Previous studies have used the AGN bolometric luminosity ($L_\mathrm{bol}$) as an indicator of black hole (BH) growth and therefore linked the quasar radio loudness to $L_\mathrm{bol}$ \citep[e.g.][]{ho_radio_2001,blundell_optically_2001,ulvestad_statistical_2001}{}. However, any observed correlation between quasar radio emission and $L_\mathrm{bol}$ suffers from the degeneracy between the black hole's Eddington-scaled accretion rate and BH mass, since both massive and highly-accreting black holes can be optically-bright. Introducing BH mass into the correlation can help break this degeneracy and uncover the leading factor that contributes to the quasar radio power, be it the BH accretion rate, BH mass, BH spin \citep[][]{wilson_difference_1995}, or other related evolutionary effects. \par

Meanwhile, understanding the role of the BH mass in the production of quasar radio emission is also a critical step in resolving the evolutionary path of radio quasars. BH masses are found to have a tight correlation with the stellar masses of their host galaxies' bulge \citep[see][and the reference therein]{kormendy_coevolution_2013}{}{}, which is now considered to be critical evidence of the co-evolution between the galaxy and the black hole through feedback processes \citep[see e.g.][]{fabian_observational_2012, heckman_coevolution_2014}{}{}. A possible connection (or not) between the BH mass and the quasar radio jet will help us determine the role of AGN jets in the feedback process, which can then serve as a building block in the galaxy evolution model. \par

Thanks to a series of optical survey campaigns, especially the Sloan Digital Sky Survey \citep[SDSS;][]{york_sloan_2000, schneider_sloan_2010, ross_sdss-iii_2012}{}{}, previous studies have established the robustness of virial black hole (BH) mass estimators \citep[e.g.][]{kaspi_reverberation_2000, vestergaard_determining_2006}, and sensible estimations of BH masses can now be derived using a set of broad-line full-width half maxima (FWHM) and the continuum luminosity from single-epoch optical spectra \citep{shen_catalog_2011,shen_sloan_2019}. However, despite the expanding sample set and improved observing methods, previous studies have failed to reach an agreement on whether the BH mass is correlated with the quasar radio loudness. Studies including \citet{mclure_relationship_2004}, \citet{seymour_massive_2007}, and \citet{whittam_mightee_2022} reported an increase in the number fraction of RL AGNs above the BH mass threshold of $10^{8-9}M_\odot$, suggesting that the RL AGNs tend to host more massive black holes than RQ AGNs, using radio observations from the NRAO VLA Sky Survey \citep[NVSS;][]{condon_nrao_1998}, the FIRST survey \citep[Faint Images of Radio Sky at Twenty Centimeters;][]{becker_first_1995}, and the MeerKAT International GHz Tiered Extragalactic Exploration \citep[MIGHTEE;][]{jarvis_meerkat_2017}, respectively.  \par

On the other hand, using recent low-frequency radio data from LOw-Frequency ARray \citep[LOFAR;][]{van_haarlem_lofar_2013}, studies including those by \citet{gurkan_lotsshetdex_2019}, \citet{macfarlane_radio_2021}, and \citet{arnaudova_exploring_2024} failed to find any correlation between BH mass and radio-loudness or differences in average BH masses between the RL and RQ population. Although using different approaches, these studies came to the same conclusion while sharing a similar data set (SDSS-detected quasars observed by LOFAR). \citet{gurkan_lotsshetdex_2019} compared the radio-loudness-$M_\textrm{BH}$ distribution for LOFAR-detected quasars and the LOFAR non-detections, and found no significant correlation between $M_\textrm{BH}$ and radio loudness. \citet{macfarlane_radio_2021} separated the quasars into hosting high-mass and low-mass BHs based on whether they were above or below the mean $M_\textrm{BH}$ of the entire sample set, before fitting their radio flux density distribution to the two-component model that separates the AGN and host galaxy contribution in the quasar radio emission, finding that the AGN component shows little difference between their high-mass and low-mass samples. \citet{arnaudova_exploring_2024} adopted a selection criterion of $R=\log_{10}(L_\textrm{1.4GHz}/L_{i})>1$ for RL quasars (where $L_\textrm{1.4GHz}$ and $L_{i}$ stand for the 1.4 GHz radio luminosity and $i$-band optical luminosity, respectively), and found no systematic difference in the distribution of BH mass estimates between RL and RQ quasars. By performing a null hypothesis test on the stacked spectra of quasars in $z-\mathcal{M}_i-M_\textrm{BH}$ space (where $\mathcal{M}_i$ is the absolute $i$-band magnitude), they also concluded that the BH mass cannot fully explain the difference between the stacked spectra of RL and RQ quasars. Therefore, they argued that BH mass is not the driving force behind quasar radio emission. \par

This disagreement with previous studies is, at least in part, caused by a lack of consensus on the various components that drive the quasar radio emission; as a result, different works adopted different definitions of radio loudness based on observational results in different frequency bands. Therefore, it is hard to integrate these different observational data sets into one physical picture. To resolve this issue, based on the two-component model from \citet{macfarlane_radio_2021}, we proposed a new Bayesian model in \citet[][hereafter \citetalias{yue_novel_2024}]{yue_novel_2024} that separates the host galaxy star formation (SF) and the contribution of AGN activity to quasar radio emission using the observed radio flux density distribution of quasars. By fitting the model against the LOFAR-observed radio flux density distribution of SDSS quasars, we were able to give a robust estimation of the mean star formation rate (SFR) of the quasar host galaxy and determine the normalisation of a power-law jet luminosity function for any subpopulation of quasars with a given absolute \emph{i}-band magnitude ($\mathcal{M}_i$) and redshift ($z$). We were able to explore how the evolution of the host galaxy SFR and AGN activity vary with $L_\mathrm{bol}$ and $z$, and also, by binning the quasars by their physical properties, to study the connection between different components of quasar radio emission and quasar colour. This approach can then be expanded to study the individual evolutionary patterns of host galaxy SFR and AGN activity with BH mass and Eddington ratio, under a physically motivated model. Therefore, we aim to answer these questions:

\begin{enumerate}
    \item Is BH mass associated with the radio loudness of quasars?
    \item Do the current definitions of quasar radio loudness correctly reflect the underlying physical processes?
    \item  Does the BH mass/accretion mode govern the AGN jet production mechanism?
\end{enumerate}

This paper is structured as follows: Section~\ref{sec:data} describes the data we used to build an optically selected quasar sample observed with LOFAR. Section~\ref{sec:result} briefly summarises the parametric model we proposed to characterise quasar radio emission and presents our result on the evolution of quasar radio emission with different BH masses. Section~\ref{sec:analysis} compares our result with those in the previous literature, which leads to a discussion on the origin of radio excess in more massive quasars. Sections \ref{sec:test} and~\ref{sec:discussion} discuss various scenarios that could explain our results. Finally, a summary of our conclusions can be found in Section~\ref{sec:conclusion}. Throughout this work, we assume a $\mathrm{\Lambda CDM}$ cosmology with parameter values published in the WMAP9 result \citep[][]{hinshaw_nine-year_2013}{}{}.

\section{Data}
\label{sec:data}

\subsection{Sloan Digital Sky Survey (SDSS) quasar sample}
\label{sec:data_sdss}

Our parent quasar sample is the SDSS Data Release 16 quasar (DR16Q) catalogue, with the photometric data presented in \citet{lyke_sloan_2020} and the associated spectroscopic data presented in \citet{wu_catalog_2022}. The DR16Q catalogue is constructed using observations from SDSS-I/II/III and IV epochs, processed with the final eBOSS SDSS reduction pipeline (v5\_13\_0), and validated spectroscopically following the criteria outlined in Table 1 of \citet{paris_sloan_2018}. It incorporates previously identified sources from the DR14Q, DR12Q, and DR7Q catalogues \citep{paris_sloan_2018,ross_sdss-iii_2012,schneider_sloan_2010}. For quasars with redshifts ranging from $0.8<z<2.2$, the identification of quasars is carried out using the decision tree detailed in \citet{dawson_sdss-iv_2016}; for quasars with redshifts $z>2.2$, the identification is based on the Ly$\alpha$ forest measurements as described in \citet{myers_sdss-iv_2015}. There are no inconsistencies observed in the characteristics (e.g., redshift, bolometric luminosity) of quasars identified through different methodologies. The total count of detected quasars exceeds $480,000$ and $239,000$ for the low- and high-redshift categories, respectively. Additional optical characteristics of quasars from the SDSS dataset, such as absolute \emph{i}-band magnitude\footnote{The \emph{i}-band magnitudes in the DR16Q catalogue were K-corrected to $z=2$, while in our dataset, they are K-corrected to $z=0$ using the original data and assuming a spectral index of 0.5 \citep{richards_sloan_2006}.}, are enriched by survey data from e.g. FIRST and the Wide-field Infrared Survey Explorer \citep[WISE;][]{wright_wide-field_2010}{}{} through cross-referencing between datasets; the details of the cross-matching process are described in \citet{wu_catalog_2022}.

\subsubsection{Black hole mass}
\label{sec:data_bhmass}

We adopt the BH mass measurements from the values presented in the spectroscopic SDSS DR16Q catalogue \citep[][]{wu_catalog_2022}{}{}. These masses are obtained by fitting the SDSS quasar spectra to a global continuum + emission line model, using the public PyQSOFIT code \citep[][]{shen_sloan_2019}{}{}. Three BH mass indicators are presented in the catalogue: \ion{H}{$\beta$}, \ion{Mg}{II}, and \ion{C}{IV}. The BH masses are estimated from single-epoch observations of these mass indicator emission lines, using the virial relation provided in \citet{schneider_sloan_2010}: 

\begin{equation}
    \log\left(\frac{M_{\textrm{BH}}}{M_{\odot}}\right)=a+b\log\left(\frac{\lambda L_\lambda}{10^{44}\textrm{erg}\:\textrm{s}^{-1}}\right)+2\log\left(\frac{\textrm{FWHM}}{\textrm{km}\:\textrm{s}^{-1}}\right),
\end{equation}

\noindent where $(a,b)=(0.910, 0.50)$ for \ion{H}{$\beta$}, $(0.740,0.62)$ for \ion{Mg}{II}, and $(0.660,0.53)$ for \ion{C}{IV}. \par

\begin{figure}     
    \includegraphics[width=\columnwidth]{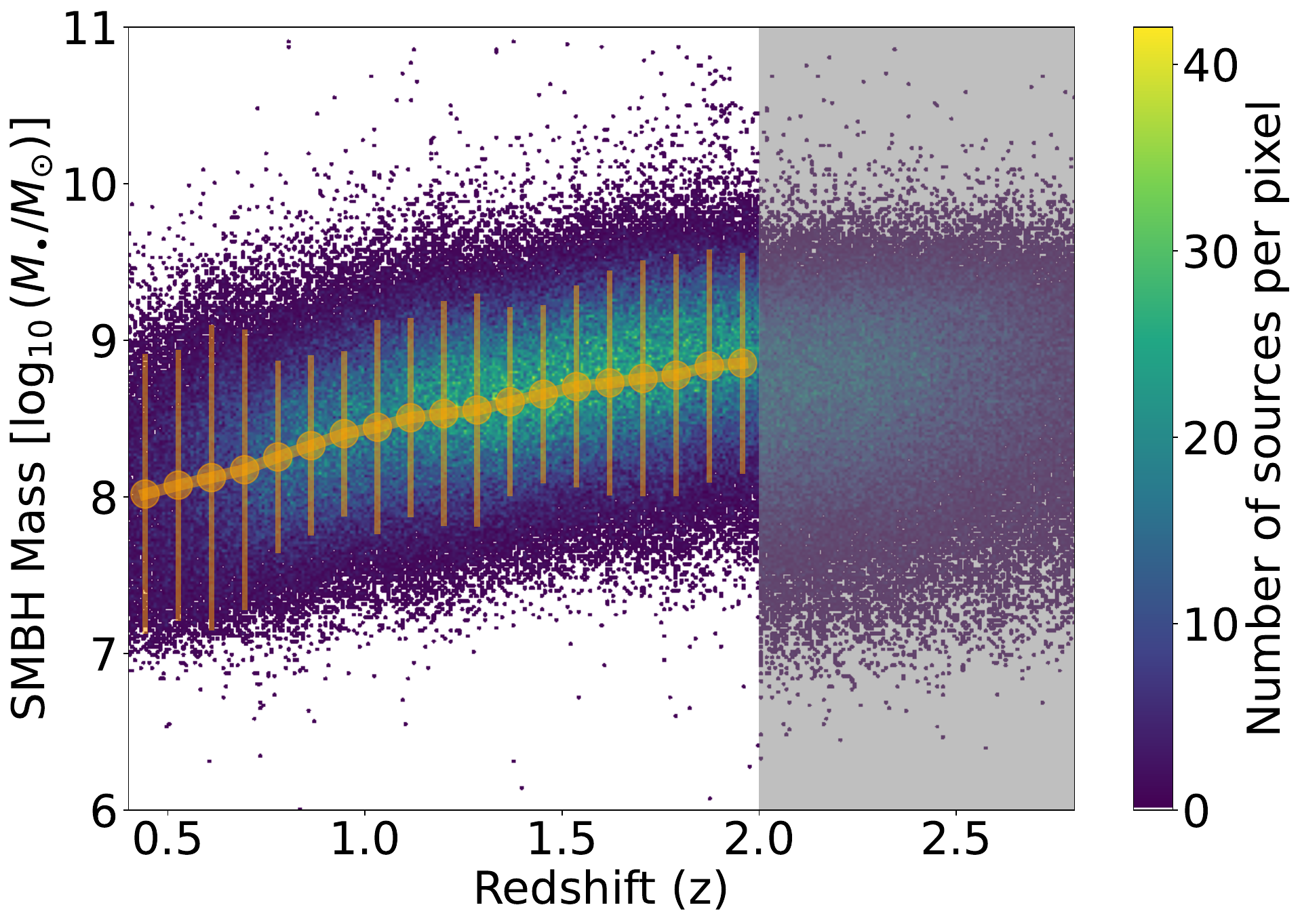}
    \caption{BH mass-redshift distribution of the sample used in this work. Yellow dots mark the average measured BH mass within each redshift bin, while the error bars show the standard deviation of BH masses within the corresponding redshift bin. Average measured BH masses lie between $M_\textrm{BH}=10^8\sim10^9M_\odot$, and maintain a good range throughout the redshift bins. The shaded area ($z>2$) marks the parameter space where only \ion{C}{IV} BH masses are available; these values show discrepancy with BH masses measured via \ion{Mg}{II} and \ion{H}{$\beta$}, possibly due to AGN outflows that introduced bias to the emission line FWHM measurement. Therefore sources with $z>2$ are excluded from our analysis.}
    \label{fig:mass_dist}
\end{figure}

We present our sample BH mass distribution with redshift up to $z=2.8$ in Figure~\ref{fig:mass_dist}. The \ion{H}{$\beta$} and \ion{Mg}{II} measurements show good consistency in the mass-redshift space, with average values (shown in light yellow dots) lying between $M_\textrm{BH}=10^8\sim10^9M_\odot$. The error bars indicate the standard deviation of BH masses within the corresponding redshift bin, from which we can conclude that the BH mass distribution maintains a good dynamical range and shows no signs of selection effects across the entire redshift range. \par

On the other hand, for sources with redshift beyond $z=2$ where only \ion{C}{IV} measurements are available, the emission line FWHM measurements appear to be discrepant, leading to a jump in the measured $M_\textrm{BH}$ values with redshift at $z=2$, as shown by the shaded area in Figure~\ref{fig:mass_dist}. Previous works, including \citet{coatman_correcting_2017}, have attempted to investigate this discrepancy and provide ways to correct the bias in these measurements. In particular, \citet{coatman_correcting_2017} argued that such discrepancy may be caused by the emission line displacement due to \ion{C}{IV} blueshift, which is in turn connected to AGN outflow activities. However, none of the correction methods appear to be conclusive, and neither do they resolve the tension between \ion{C}{IV} and \ion{Mg}{II} $M_\textrm{BH}$ measurements. In addition, other studies have shown a strong correlation between AGN outflow and enhanced radio emission from AGN \citep[e.g.][]{morabito_origin_2019,petley_connecting_2022}, suggesting that biased mass estimations from \ion{C}{iv} emission line are associated with quasar radio excess at $z>2$. As a result, to avoid any biases, we exclude the sources with $z>2$ and restrict our samples to lower redshift ($z\leq2$) where the \ion{Mg}{ii} and \ion{H}{$\beta$} estimates are available. \par

\subsection{Matching LoTSS data with SDSS quasars}

\subsubsection{LOFAR Two-metre Sky Survey (LoTSS)}
\label{sec:data_lotss}

\begin{figure}     
    \includegraphics[width=\columnwidth]{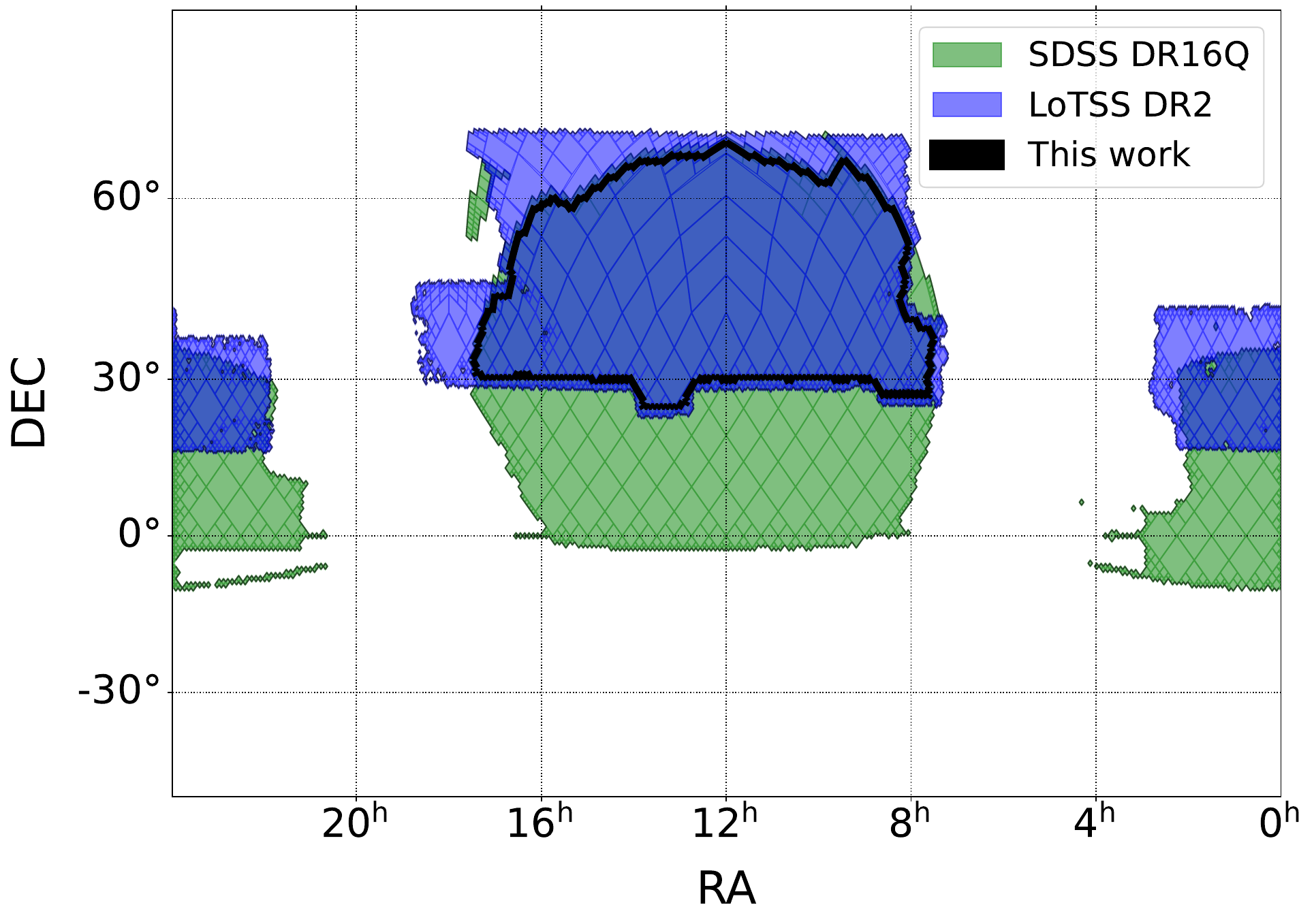}
    \caption{A comparison of the sky coverages from LoTSS DR2 (blue) and SDSS DR16Q (green) survey catalogues. The black line encompasses the sky area studied in this paper, which is the overlap between the LoTSS DR2 and SDSS DR16Q survey areas in the north galactic cap.}
    \label{fig:lotssdr2sky}
\end{figure}

As in \citetalias{yue_novel_2024}, we obtain the radio measurements for our quasar samples via the LOFAR Two-metre Sky Survey \citep[LoTSS;][]{shimwell_lofar_2017, shimwell_lofar_2019, shimwell_lofar_2022}, which is a wide-field imaging survey conducted with the LOFAR HBA (high-band antenna) system. It aims to survey the entire northern sky in the 120-168 MHz radio band, with a desired rms noise level of approximately $100\mu\mathrm{Jy\ beam}^{-1}$, an angular resolution of $6 \arcsec$, and a positional accuracy better than $0.2 \arcsec$. The most comprehensive catalogue to date is the LoTSS DR2 catalogue presented by \citet{shimwell_lofar_2022}, which covers an area of 5,720 $\mathrm{deg}^2$ of the sky and achieves a median rms noise of $83\mu\mathrm{Jy\ beam}^{-1}$. The current LoTSS DR2 catalogue comprises approximately 4,400,000 radio-detected sources. The sky coverage, in comparison with the SDSS DR16 quasar catalogue, is illustrated in Fig.~\ref{fig:lotssdr2sky}. A significant portion of the sky area in LoTSS DR2 overlaps with the SDSS DR16 quasar catalogue, making it an excellent resource for multi-wavelength studies. \par

In the LoTSS DR2 catalogue, radio sources were extracted from LoTSS images using PyBDSF, the Python Blob Detector and Source Finder \citep{mohan_pybdsf_2015}. These sources were pinpointed based on the peak radio flux densities surpassing the $5\sigma$ threshold in the LoTSS DR2 images. Although PyBDSF is proficient at detecting regions of radio emission, it does not always properly group them to physical sources. To mitigate misassociations by PyBDSF, a combination of statistical methods and thorough visual examination, facilitated by the LOFAR Galaxy Zoo, was employed to validate that the radio catalogue reflects the genuine distribution of radio sources \citep[][]{hardcastle_lofar_2023}{}{}. In addition to the revised radio catalogue, \citet{hardcastle_lofar_2023} also detail the process of cross-referencing the LoTSS DR2 sources with the optical-infrared counterparts of the WISE \citep{wright_wide-field_2010} and DESI Legacy Imaging \citep{dey_overview_2019} surveys, employing a methodology similar to that outlined in \citet{williams_lofar_2019} and \citet{kondapally_lofar_2021}.\par

\subsubsection{Radio Flux Densities from LoTSS}
We obtain the radio flux density measurements for SDSS DR16Q quasars from the LoTSS DR2 catalogue and the original LoTSS mosaic. While the details of the process are described in \citetalias{yue_novel_2024}, we briefly summarise our approach as follows: \par

\begin{enumerate}
    \item For sources with radio flux densities that surpass the detection limit of $5\sigma$ in LoTSS DR2, we cross-matched the positions of quasars in the SDSS catalogue with those in the source-associated LoTSS catalogue using a matching radius of $1.5\arcsec$. Note that whenever feasible, we used the coordinates from the cross-referenced optical catalogues of \citet{hardcastle_lofar_2023} for LoTSS positions. \par
    
    \item For sources with radio flux densities below the LoTSS $5\sigma$ detection limit and thus not recorded in the LoTSS DR2 catalogue, we performed forced photometry on LoTSS mosaics using the method described in \citet{gloudemans_low_2021}. This approach is motivated by the assumption that most quasars are unresolved in LoTSS DR2 images, while the astrometric uncertainty in LoTSS DR2 ($\lesssim0.2\arcsec$) is significantly smaller than the LoTSS pixel scale ($1.5\arcsec$). The extracted radio flux density is considered to be the highest pixel value within a $3\mathrm{px}\times 3\mathrm{px}$ aperture centred on the source position in the SDSS catalogue. The uncertainty in the flux density is calculated based on the standard deviation of pixel values in a $100\mathrm{px}\times 100\mathrm{px}$ cut-out region surrounding the central pixel. As discussed in \citetalias{yue_novel_2024}, although these individual measurements are rather noisy, they retain valuable information about the distribution of radio flux densities of the population as a whole.
\par
\end{enumerate}

\subsection{Building a LoTSS-SDSS quasar sample}
\label{sec:data_build}

\begin{figure}     
    \includegraphics[width=\columnwidth]{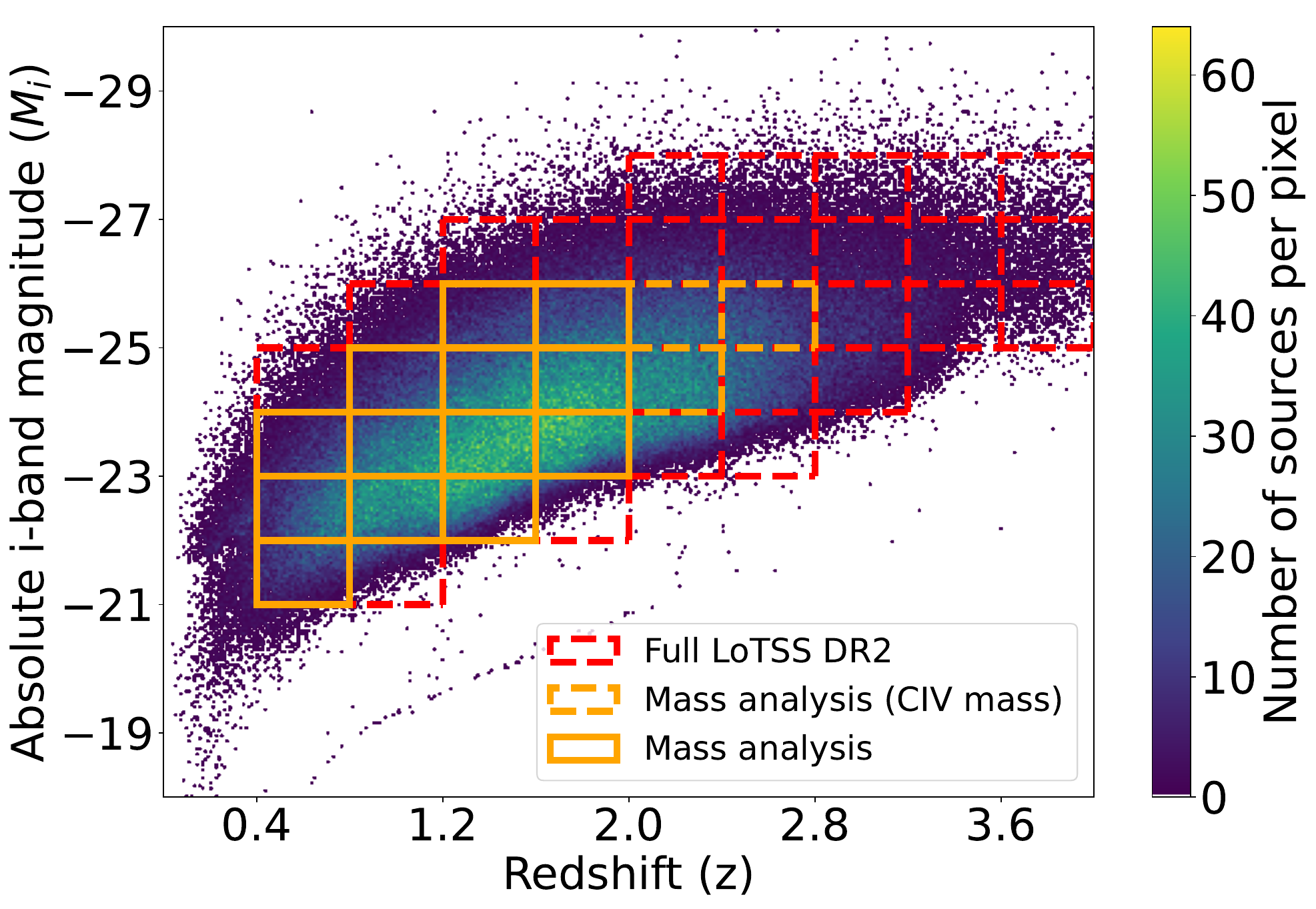}
    \caption{Coverage of parameter space in $\mathcal{M}_i-z$ plane. The red dashed grids show the total LoTSS DR2-SDSS sample set included in \citetalias{yue_novel_2024}, while the solid orange grids show the samples used in this work: each grid hosts at least 5,000 quasars with estimated BH masses from either \ion{H}{$\beta$} or \ion{Mg}{II} measurements. The dashed orange grid cells each host more than 5,000 quasar samples but have only \ion{C}{IV} BH mass estimations. These grid cells are excluded from this work but may be reintroduced into the sample once high-quality spectroscopic data becomes available in future surveys including WEAVE. }
    \label{fig:opticalproperty}
\end{figure}

We build our quasar sample set based on the method presented in \citetalias{yue_novel_2024} by extracting LoTSS radio flux density measurements for the parent SDSS DR16Q quasar samples. As described in \citetalias{yue_novel_2024}, we have removed the following sources from the SDSS DR16Q sample set:

\begin{enumerate}
    \item Overly luminous sources (with absolute \emph{i}-band magnitude brighter than $\sim-40$) that are likely artefacts of the SDSS data reduction pipeline; \par 

    \item Sources in the 0h field (as illustrated in Figure~\ref{fig:lotssdr2sky}) that exhibit consistently higher uncertainties in radio flux density compared to those in the 13h region, and quasars that are not covered by LoTSS being located either outside the target field or within the spaces between the LoTSS mosaics. \par

    \item Sources in the SDSS DR16Q catalogue for which the only selection criteria for spectropscopic observation was the radio properties. \par

    In addition to the restrictions above, we have applied the following additional constraints for the specific analysis in this paper: \par

    \item Given that our fitting approach uses sources within each grid in the $\mathcal{M}_i$—$z$ plane (see Figure~\ref{fig:opticalproperty}), we focus only on cells within the $\mathcal{M}_i$—$z$ grids that contain more than 5,000 sources. This is because during the mass dependence analysis, each $\mathcal{M}_i$—$z$ grid need to be binned into 5 sub-grids by the black hole mass percentiles within each grid, and the minimum sample size required by our fitting approach is $\sim1,000$ \citep[see][]{yue_novel_2024}{}{}. \par

    \item In \citet{wu_catalog_2022}, the final black hole masses are adopted from estimations from different mass indicators based on redshift range (\ion{H}{$\beta$} for $z<0.7$, \ion{Mg}{II} for $0.7\leq z\leq2.0$).  We discard the sources where only \ion{C}{IV} measurements are available due to the reasons listed in Section~\ref{sec:data_bhmass}.
\end{enumerate}

The final sample consists of 222,782 quasars. The samples are characterised by their distribution in the $\mathcal{M}_i-z$ plane, as shown in Fig.~\ref{fig:opticalproperty}. While the sample is naturally biased towards brighter absolute magnitudes at higher redshifts, it still maintains a good sampling range across the entire parameter space. \par

\section{The black hole mass impact on quasar radio continuum emissions}
\label{sec:result}

\subsection{Modelling the quasar radio flux density distribution}

In \citetalias{yue_novel_2024}, we presented a fully Bayesian two-component model that disentangles the host galaxy and the AGN contribution in a given quasar radio flux density distribution. This model can then be adopted to analyse the dependence of the quasar host galaxy and AGN radio emission on black hole mass independently. Here we briefly summarise the structure of our Bayesian model. \par

In order to translate the observed quasar radio flux density distribution to a parametric statistical function, we followed the methodology developed in \citet{roseboom_cosmic_2014} that allows us to stack the sources and analyse them as a whole. More specifically, our model calculates the probability $P(s_j)$ of each input source $j$ within a given $\mathcal{M}_i-z$ grid (as shown in Figure~\ref{fig:opticalproperty}) having an observed flux density $s_j$. Through Bayes' theorem, the likelihood of the fitted model $\boldsymbol{\mu}$ is defined as

\begin{equation}
    P(\boldsymbol{\mu}|\{s_j\})\propto P(\boldsymbol{\mu})P(\{s_j\}|\boldsymbol{\mu})=P(\boldsymbol{\mu}) \prod_j P(s_j|\boldsymbol{\mu}). 
\end{equation}\label{eq:bayesian}

\noindent where $\{s_j\}$ is the distribution of the observed radio flux density within the $\mathcal{M}_i-z$ grid. \par

We then derive the expression of our model probability density function (PDF) $P(s_j|\boldsymbol{\mu})$ based on the two-component assumption presented in \citet{macfarlane_radio_2021} and \citetalias{yue_novel_2024}, where \emph{every} quasar is thought to host radio emission from both host galaxy SF and AGN activity. Assuming that quasars within a $\mathcal{M}_i-z$ grid share similar physical properties, we model the SF component with a log-Gaussian PDF centred at the 150 MHz radio luminosity $\log(L_\mu/[\mathrm{W~Hz}^{-1}])$ that traces the typical star-formation rate $\Psi$ within the grid,  and with a scatter of $\sigma_\mu/\textrm{dex}$: 

\begin{equation} \label{eq:L_SF}
    P_\mathrm{SF}(L)dL=\left[\frac{1}{\sigma_\mu\sqrt{2\pi}}\exp\left(\frac{\log L-\log L_\mu}{2\sigma_\mu^2}\right)\right]\frac{dL}{L}.
\end{equation}

The fitted values of $L_\mu$ and $\sigma_\mu$ can then be converted to SFR ($\Psi$) and scatter in SFR ($\sigma_\Psi$) using the mass-independent SFR-radio luminosity correlation in \citet{smith_lofar_2021}. We model the AGN component by extrapolating the luminosity function of radio-loud AGNs to lower luminosities, assuming that both radio-loud and radio-quiet AGNs share the same mechanism for generating radio emissions, only with different power efficiency. The AGN component is therefore characterised by a power-law PDF with a slope of $\gamma$ and a normalisation parameter $\phi$: 

\begin{equation} \label{eq:L_AGN}
    P_\textrm{AGN}(L)dL\propto\phi L^{-\gamma}dL, 
\end{equation} 

\noindent where $\phi$ can be translated into the radio-loud fraction $f$, defined in \citet{macfarlane_radio_2021} as the fraction of sources with 150 MHz luminosity of the AGN component above $10^{26}\mathrm{W/Hz}$ . To maintain consistency with \citet{macfarlane_radio_2021} and \citetalias{yue_novel_2024},  we characterise our best fit model with the set of parameters $\{\Psi,\sigma_\Psi,\gamma,f\}$.

\begin{figure*}     
    \includegraphics[width=\textwidth]{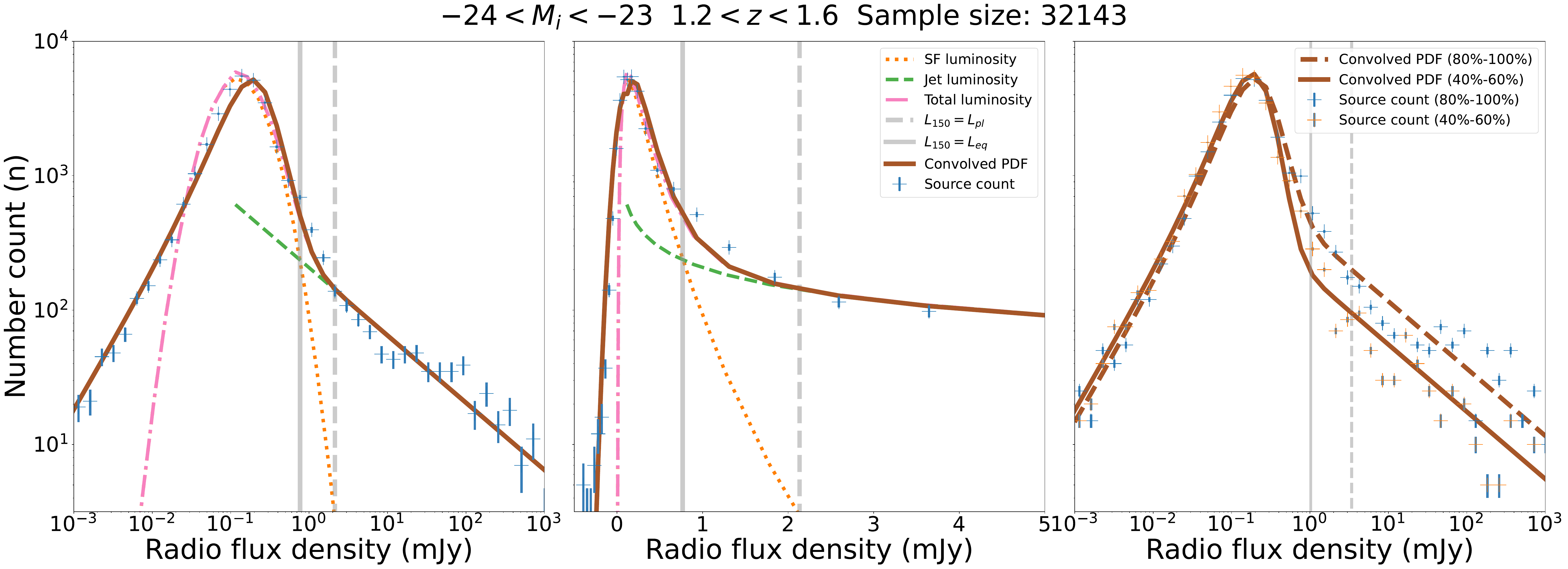}
    \caption{Radio flux density distribution and the corresponding model best-fit in one of the representative $\mathcal{M}_i-z$ bins explored in this study. The left and middle panels show the distribution and fitted relation for the entire quasar population within the grid, in log and linear scales respectively. The orange dotted line and green dashed line represent the SF component from the host galaxy activity (log Gaussian) and the jet component from the AGN activity (single power-law) respectively. The pink dashed-dotted line shows the combined PDF of the two-component radio flux density distribution model, and the brown solid line shows the probability density function (PDF) after convolving with the observational error. This convolution spreads the original PDF at the faint end out to negative flux density values, leading to an offset between the convolved and original PDFs at flux densities lower than $100\mathrm{\mu Jy}$. The distribution of measured radio flux densities is shown as blue crosses, indicating a good match between the actual distribution and the best-fit model result. The vertical solid and dashed grey lines mark the criteria proposed in this work for identifying host galaxy- and AGN-dominated quasars, which will be discussed in section~\ref{sec:analysis}. In the right panel, we present our fitting results for medium BH mass (with values in the 40\%-60\% of the mass distribution) and high BH mass bins (80\%-100\%) in solid and dashed lines, respectively. The measured radio flux density distributions of both bins are mapped against the best-fits as orange and blue crosses.}
    \label{fig:pdf}
\end{figure*}

The left and middle panels of Figure~\ref{fig:pdf} show the example probability density function within one of the $\mathcal{M}_i-z$ grids that we explored in this work, in log and linear space, respectively. The orange dotted line shows the fitted underlying radio flux density distribution of host galaxy SF activity, while the green dashed line shows the underlying distribution of AGN activity. Combining these two components, we have the total underlying radio flux density distribution for the quasar population (pink dash-dotted line). Finally, the PDF convolved with the radio flux density error is shown in the solid brown line. Our convolved PDF shows good agreement with the observed quasar radio flux densities (blue crosses) across all $\mathcal{M}_i-z$ grids, indicating that our model provides a good description of the quasar sample. \par

\subsection{The dependence of quasar radio emission on BH mass}
With our model being able to separate SF and AGN features in a radio flux density distribution containing at least 1,000 sources (\citetalias{yue_novel_2024}), we can take an $\mathcal{M}_i-z$ grid cell and further bin it down to sub-samples by their measured black hole masses. By comparing the fitting result across different sub-samples within one $\mathcal{M}_i-z$ grid cell, we can investigate how the SF and AGN components in quasar radio emission vary separately with black hole masses. With the average uncertainties of our black hole masses being around $0.1~\textrm{dex}$ and the typical range of masses within a bin being between $\log(M/M_\odot)=7$ and $10$, we bin the black hole masses within a $\mathcal{M}_i-z$ grid cell into 5 sub-sample sets, each containing sources with measured black hole mass in quintiles $0\%-20\%$, $20\%-40\%$, $40\%-60\%$, $60\%-80\%$, and $80\%-100\%$ based on the mass distribution in the $\mathcal{M}_i-z$ grid cell\footnote{We pick this binning method to balance between the ability to see a quantified trend and the possibility of cross-contamination between different bins due to uncertainties in BH mass measurement. The typical width of the BH mass bin is $\sim0.25~\textrm{dex}$; while this width is larger than the quoted statistical uncertainties of the BH mass measurements, several studies \citep[e.g.][]{mclure_measuring_2002,shen_mass_2013,peterson_measuring_2014,chaves-montero_black_2022,shen_sloan_2024} indicate that the typical scatter between BH masses obtained from single-epoch observations and reverberation campaigns is $0.3\sim0.5~\textrm{dex}$, which is larger than (or similar to) our typical bin widths. There may therefore be some cross-contamination between the bins, which would weaken any observed trend.}. The right panel in Figure~\ref{fig:pdf} takes the third and fifth quintile BH mass bins as an example to show how different best-fit model parameters affect the final convolved PDFs and that the parameterisation we use gives an accurate reflection of the observed radio flux density distributions within each sub-sample. \par

\begin{figure*}     
    \includegraphics[width=\textwidth]{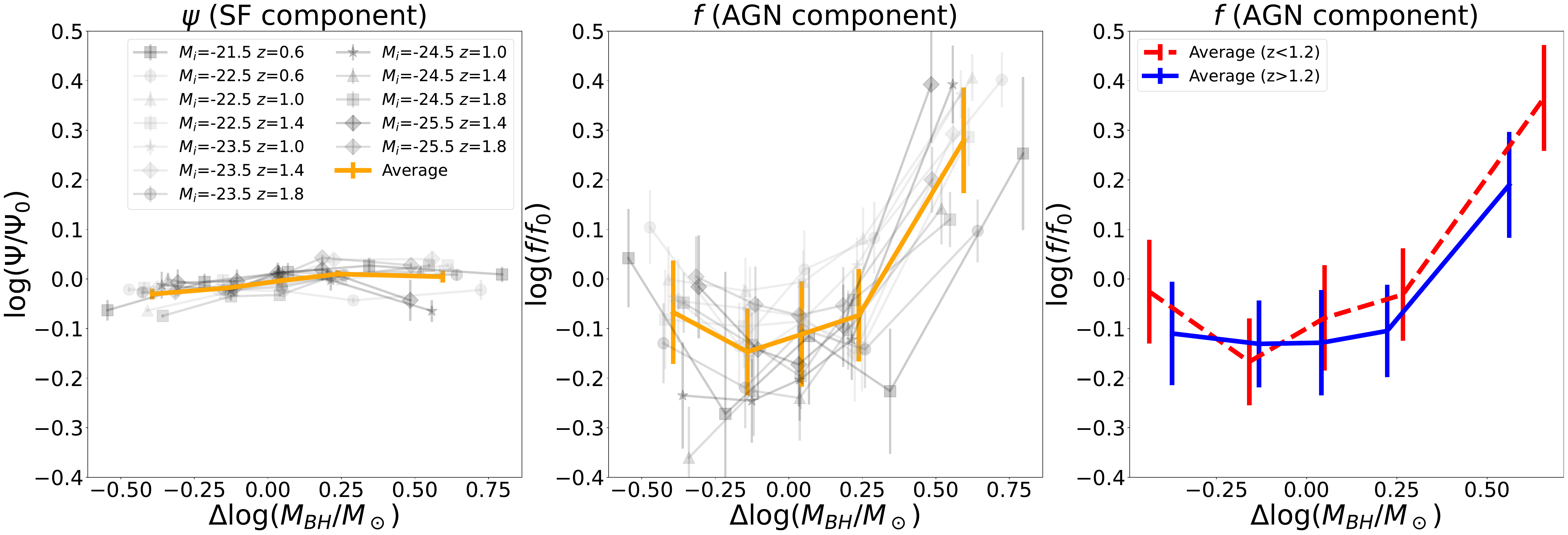}
    \caption{The evolution of SF and AGN contributions to the quasar radio flux density distribution with BH mass, as measured from the model best-fits using quasar samples split by BH mass percentiles. Within each $\mathcal{M}_i-z$ grid cell, the BH masses are divided into 5 quintiles (0\% - 20\%, 20\% - 40\%, 40\% - 60\%, 60\% - 80\%, 80\% - 100\%) and are characterised by the difference between the average BH mass in each percentile bin and the average from the entire $\mathcal{M}_i-z$ grid cell ($\Delta\log M_\textrm{BH}$). The SF and AGN contributions are characterised by the relative variation of the model parameters ($\Psi/\Psi_0$, $f/f_0$), where $\Psi$, $f$ are the best-fit parameter values within each mass quintile and $\Psi_0$, $f_0$ are the best-fit value using quasars from the entire $\mathcal{M}_i$-$z$ grid cell. The grey lines show the correlation in each $\mathcal{M}_i-z$ bin, and the yellow line presents the correlation averaged over the entire $\mathcal{M}_i-z$ parameter space. The SF contribution shows little variance with increasing BH mass (left panel). Similarly, the AGN contribution within most of the BH mass range shows little variance over most of the BH mass range, but the most massive quasars (top 20\% of the mass percentile) show enhanced radio emission from AGN activities (middle panel), which confirms the pattern seen in Figure~\ref{fig:pdf}. The right panel shows the variation of AGN activity split into low ($z<1.2$) and high ($z>1.2$) redshift ranges, which confirms that this trend is uniformly seen across all redshifts.}
    \label{fig:result_mass}
\end{figure*}

Going into more detail, Figure~\ref{fig:result_mass} shows the continuous evolution of our fitted parameters with black hole mass percentiles in all of the $\mathcal{M}_i-z$ grid cells included in this work. Here we focus on the parameters $\Psi$ and $f$, as they are key indicators of the levels of host galaxy SF activity and AGN activity, respectively. The parameter $\gamma$ is fixed at $\gamma=1.5$ based on the result from extrapolating the bright end of the distribution function as described in \citetalias{yue_novel_2024}, while $\sigma_\mu$ remains a free parameter in our fit. The inferred $\sigma_\mu$ values are typically between 0.2 and 0.3, with less than 5\% variation with BH mass; thus, they are excluded from the discussion here. The sub samples are characterised by the deviation between the average black hole mass within the sub sample set and the average black hole mass within the $\mathcal{M}_i-z$ grid cell, denoted by $\Delta\log M_\textrm{BH}$ in Figure~\ref{fig:result_mass}. Meanwhile, the evolution of quasar radio emission from host galaxy and AGN, i.e. our \emph{y}-axis, is characterised by the deviation between the fitted value of the parameters in the sub sample set and the fitted value in the full sample within that $\mathcal{M}_i-z$ grid cell, denoted as $\log_{10} (\Psi/\Psi_0)$ and $\log_{10}(f/f_0)$ for the SF component and the AGN component, respectively. \par

As seen in Figure~\ref{fig:result_mass}, there is little sign of variation with the black hole mass for the host galaxy SF activity across the entire sample (left panel) and for AGN activity with most of the black hole mass bins (middle panel). However, quasars with the most massive BHs (top 20\% of the BH mass distribution) are 2 to 3 times more likely to be radio-bright when compared to the quasars of the same redshift and bolometric luminosity while hosting smaller BHs, due to an enhanced AGN activity (a $\sim0.4\ \mathrm{dex}$ increase in the fitted value of $\log(f)$). We have also split the variation of AGN activity into low-redshift ($z<1.2$) and high-redshift ($z>1.2$) regimes, and showed that the trend of the variation remains consistent in all redshifts (right panel). Note that while the modelling assumes a fixed value of $\gamma$, inspection of individual fits shows that this assumption appears reasonable (e.g. the right panel in Figure~\ref{fig:pdf}). This means that the enhancement occurs across a broad range of AGN jet powers. \par

\section{A coherent picture of quasar radio loudness and black hole mass}
\label{sec:analysis}

In the above section, we have split the sample by BH mass and explored its effect on the quasar radio continuum emission. Although the relation between quasar radio luminosity and black hole mass has been visited many times in the previous literature, most of those works split the quasars by their radio properties (e.g. radio loudness) and examined the resulting BH mass distribution. While the model best-fits in Figure~\ref{fig:result_mass} outline the evolution of radio emission within continuous mass percentiles, in this section we will divide the samples by their radio brightness instead and examine their respective BH mass distributions. We can therefore bring the detected radio enhancement into the context of the overall radio luminosity and radio brightness of quasars to form a coherent picture and unify the previous literature. \par

As discussed previously, studies including \citet{mclure_relationship_2004} and \citet{whittam_mightee_2022} are in favour of the hypothesis where BH mass is connected to radio loudness due to RL quasars in their sample typically hosting more massive black holes ($M_\textrm{BH}\gtrsim10^8M_\odot$). However, other works including \citet{gurkan_lotsshetdex_2019} and \citet{arnaudova_exploring_2024} have found no correlation between radio loudness and BH mass, nor any difference between the RL and RQ quasar population in their BH mass distribution, therefore arguing that the BH mass is not a primary driving factor behind the radio-loudness of quasars. The main difference between these works is the radio loudness criteria that they used to separate the RL and RQ quasars. For studies such as \citet{mclure_relationship_2004} that use high-frequency radio data, the widely adopted definition is $R_\mathrm{flux}=f_\textrm{5GHz}/f_\textrm{4400\AA}$, where $R_\mathrm{flux}$ is the radio loudness, and $f_\textrm{5GHz}$ and $f_\textrm{4400\AA}$ are the observed fluxes from AGN in radio and optical bands, respectively. The threshold for RL/RQ quasar separation is $R_\mathrm{flux}=10$, with $R_\mathrm{flux}>10$ quasars classified as RL and the rest as RQ. More recent studies using LOFAR data \citep[e.g.][]{gurkan_lotsshetdex_2019,arnaudova_exploring_2024} used the definition based on radio luminosities: $R_\mathrm{lum}=\log_{10}(L_{1.4\textrm{GHz}}/L_i)$, where $L_{1.4\textrm{GHz}}$ is the 1.4GHz radio luminosity converted from LOFAR 150 MHz radio measurements assuming a radio slope of $\alpha=0.7$ \footnote{Here $\alpha$ is defined with $L_\nu\propto\nu^{-\alpha}$}, and $L_i$ is the luminosity observed by the SDSS $i$-band filter; the RL/RQ threshold is defined as $R_\mathrm{lum}=1$ \citep[e.g.][]{balokovic_disclosing_2012,arnaudova_exploring_2024}. \citet{gurkan_lotsshetdex_2019} performed a correlation analysis between $R_\mathrm{lum}$ and $M_\mathrm{BH}$ where no distinction was made between the RL and RQ population; however, they still characterised their quasar population with the peak values in the $R_\mathrm{lum}$ distribution ($\log_{10}(L_{150\textrm{MHz}}/L_i)\sim1.6$). Note that while both are commonly referred to as `$R$' in the literature and we sometimes use $R$ when discussing radio loudness in general, in this section we distinguish the two definitions ($R_\mathrm{flux}$ and $R_\mathrm{lum}$) with different subscripts for clarity. Alternative definitions based on radio excess are used in works including \citet{whittam_mightee_2022}, where RL quasars are defined as having radio emission above the $2\sigma$ threshold of the infrared-radio correlation (IRRC). This threshold implies that radio emission cannot be produced from SF alone for sources classified as RL. \par

However, definitions using radio flux or luminosity ratios are generally based on thresholds or boundaries in observational quantities rather than physical models. They group quasars over a wide range of redshifts and luminosities together to provide a single threshold, whereas the quasar radio emission is known to scale (non-linearly) with both of these parameters \citepalias[e.g.][]{yue_novel_2024}. Therefore, in this work, based on the two-component model presented in Section~\ref{sec:result}, we propose a new physically motivated quasar classification derived from the fitted contribution of the SF and AGN components in the quasar radio emission. We define the expression for the threshold radio luminosity $L_\textrm{eq}$ (shown as the vertical solid line in Figure~\ref{fig:pdf}), where $P_\textrm{SF}(L_\textrm{eq})dL=P_\textrm{AGN}(L_\textrm{eq})dL$, using the expressions for $P_\textrm{SF}(L)dL$ and $P_\textrm{AGN}(L)dL$ presented in Equations~\ref{eq:L_SF} and~\ref{eq:L_AGN} (note that the expressions of $P_\textrm{SF}(L)dL$ and $P_\textrm{AGN}(L)dL$ depend on the model best-fits and therefore values of $L_\textrm{eq}$ vary between different grid cells). By definition, the AGN contribution to radio emission in sources with radio luminosities below $L_\textrm{eq}$ is very likely to be smaller than the SF contribution ($P_\textrm{AGN}<P_\textrm{SF}$), and therefore these sources can be classified as SF-dominated (traditionally RQ) quasars. \par

On the other hand, as shown by the vertical dashed line in Figure~\ref{fig:pdf}, we use $L_\textrm{pl}$ as a radio luminosity lower limit for AGN-dominated sources (traditionally RL), since their fitted PDFs of radio flux density distribution are dominated by the single power law AGN component. $L_\textrm{pl}$ is defined as the luminosity threshold where the power-law component strongly dominates the PDF: $P_\textrm{AGN}(L>L_\textrm{pl})dL>0.95~P(L>L_\textrm{pl})dL$ so that 95\% of . This is a conservative approach to minimise the contamination of this AGN-dominated sample by radio intermediate (RI) sources lying at the bright end of the Gaussian SF distribution, as their radio emissions could be dominated by either SF or AGN. As a result, RI sources have radio luminosities between $L_\textrm{eq}$ and $L_\textrm{pl}$. Since the origin of radio emission in RI quasars is indistinguishable by the model, we exclude them from the following analysis. To avoid any confusion, since our quasar classification is based on actual contributions of SF and AGN activities rather than radio-loudness thresholds, we will use the terms `SF-dominated' and `AGN-dominated' instead of RQ / RL in the following analysis. \par

\begin{figure*}     
    \includegraphics[width=0.8\textwidth]{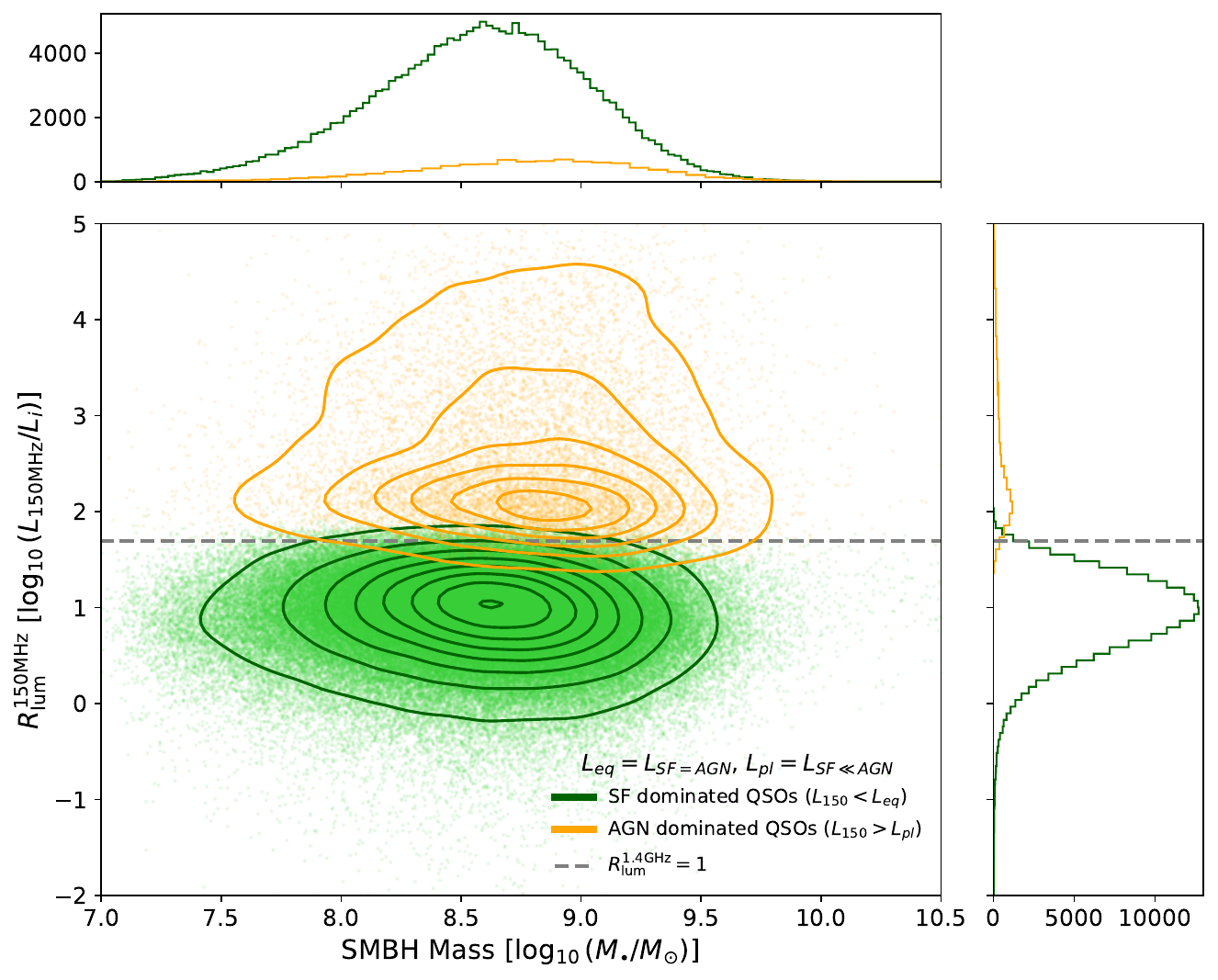}
    \caption{The distribution of AGN-dominated (light orange) and SF-dominated (dark green) quasars under our new model-driven classification in the $R_\mathrm{lum}-M_\textrm{BH}$ space, where $R_\mathrm{lum}$ is the radio loudness defined as $R_\mathrm{lum}=\log_{10}(L_{150\textrm{MHz}}/L_i)$. With characteristic luminosity defined as $L_\textrm{eq}$ where $P_\textrm{SF}(L_\textrm{eq})dL=P_\textrm{AGN}(L_\textrm{eq})dL$, SF-dominated (radio-faint) quasars are defined as quasars with radio luminosities $L<L_\textrm{eq}$, and AGN-dominated (radio-bright) quasars are defined as having radio luminosities $L>L_\textrm{pl}$, where the power-law component $P_\textrm{AGN}(L)dL$ dominates over 95\% of the total PDF when $L>L_\textrm{pl}$. The contours track the same fractions of the AGN-dominated and SF-dominated quasar populations in each class, while the histograms show marginalised distributions in the $R$ and $M_\textrm{BH}$ spaces. The histogram at the top shows the marginalised distribution of $M_\mathrm{BH}$ for AGN-dominated (light orange) and SF-dominated (dark green) quasars, and the histogram on the right shows the marginalised distribution of $R$ following the same colour code for the SF- and AGN-dominated population. Note that the bin widths of the histograms for SF- and AGN- dominated populations are defined with Scott's rule and are thus different from each other. The criterion for RQ quasars used in \citet{balokovic_disclosing_2012} and \citet{arnaudova_exploring_2024} ($R_\mathrm{lum}<1$) is shown as the horizontal grey dashed line in the right-hand histogram and the scatter plot.}
    \label{fig:scatter}
\end{figure*}

Figure~\ref{fig:scatter} shows the distribution of the SF-dominated (dark green) and AGN-dominated (light orange) quasars under our new definition, plotted in the radio-loudness-BH-mass space, where the radio loudness is defined as $R=\log_{10}(L_{150\textrm{MHz}}/L_i)$ in order to compare with previous studies. The contours highlight the number density in the distribution by marking the same percentiles for both populations. As shown by the contours, both the RL and the RQ quasar population span across the entire BH mass range, suggesting that there is no sharp bimodality in the BH mass distribution of the RL and RQ quasars. This being said, RL quasars under our classification tend to have a higher BH mass on average, as suggested by the histogram on the top panel. \par

The marginalised distributions of both populations in radio loudness and in BH mass space are shown as histograms on the side and top panels. As seen from the dashed line on the right histogram (which illustrates $R_\mathrm{lum}=\log_{10}\left(L_\mathrm{1.4GHz}/L_i\right)=1$), our definition is comparable to the criteria used in the previous literature, since the dashed line falls on the tails of both the SF- and AGN-dominated population. However, unlike previous definitions, the SF- and AGN-dominated quasar populations overlap in their radio loudness values (see the histogram on the right) due to their different levels of SF and AGN activity in the $\mathcal{M}_i-z$ space. This addresses the difference between a more physically based classification and the traditional RL/RQ quasar classification based solely on the radio loudness. \par

\begin{figure*}     
    \includegraphics[width=\textwidth]{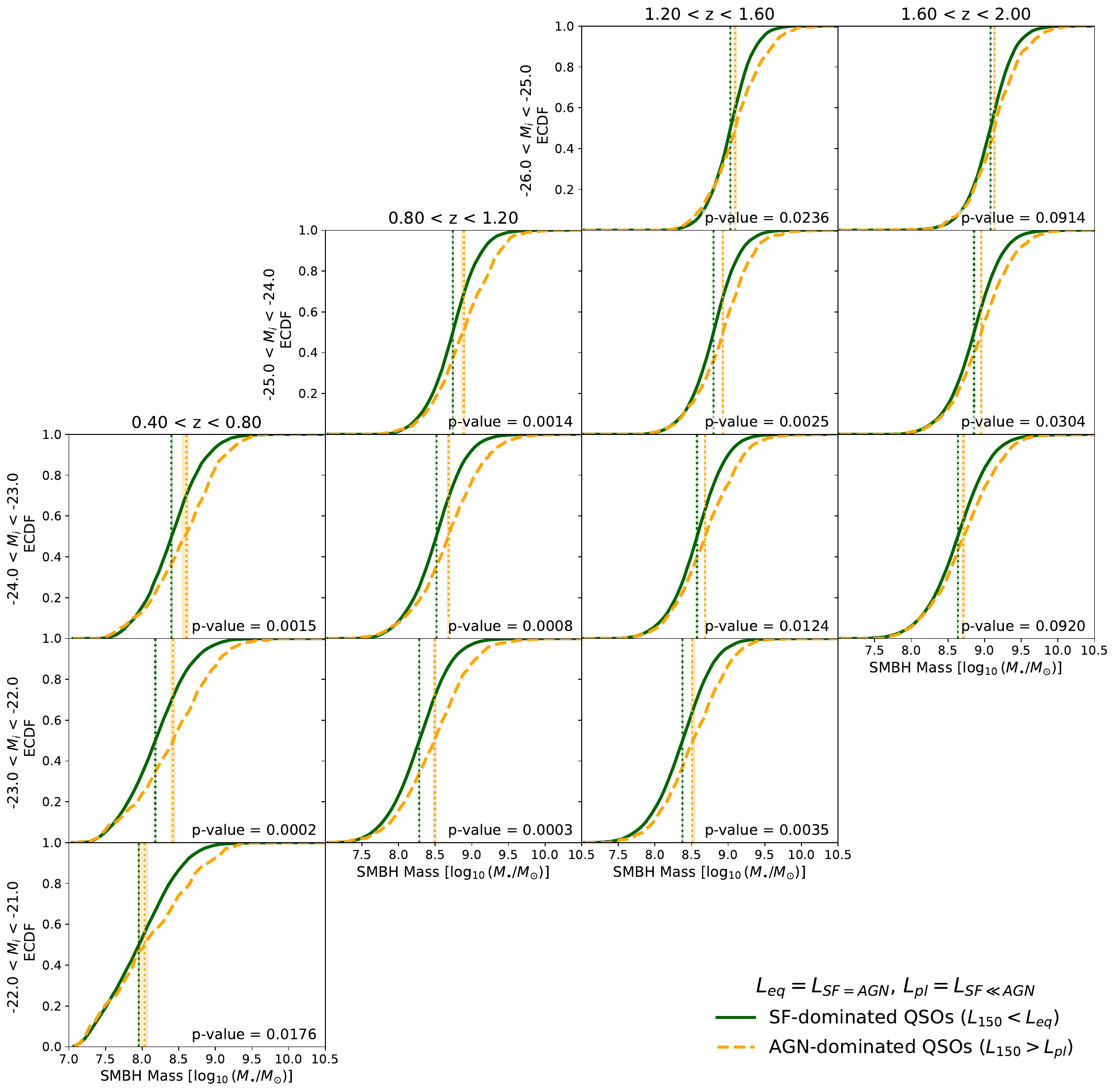}
    \caption{Cumulative mass distributions of AGN- (light orange) and SF-dominated (dark green) quasars defined in this work within each $\mathcal{M}_i-z$ grid. The grid cells are arranged by their average \emph{i} band luminosity and redshift - grid cells with fainter luminosity are placed towards the bottom, and those with lower redshift are placed towards the left. The vertical solid and dashed lines show the median BH mass for SF-dominated and AGN-dominated quasars, respectively, and the uncertainty is determined from the 16\% and 84\% confidence intervals when bootstrap sampling the distributions. We quantify the similarity between two populations with p-values calculated from K-S tests: p-values less than 0.05 reject the null hypothesis that two distributions are identical. Our new definition leads to a universal difference in the BH mass distribution between radio-faint and radio-bright quasars, suggesting that AGN-dominated quasars tend to be more massive than SF-dominated quasars.}
    \label{fig:ecdf_mass_with_fit}
\end{figure*}

In Figure~\ref{fig:ecdf_mass_with_fit}, we plot the cumulative distribution function (CDF) of the BH mass for SF- and AGN-dominated quasars under the new definition we proposed, across the explored $\mathcal{M}_i-z$ grid cells. The corresponding characteristic luminosities $L_\textrm{eq}$ and $L_\textrm{pl}$ for each grid cell are shown in Table~\ref{tab:lum_values} of Appendix~\ref{sec:appendix_result} as a reference. Within each grid cell, the black hole mass CDF of SF-dominated quasars is shown as a green solid line, whereas for AGN-dominated quasars, the CDF is shown as an orange dashed line. Both distributions are characterised by their median black hole masses, shown by the green and orange vertical lines respectively. Across many $\mathcal{M}_i-z$ grid cells, the cumulative distributions of black hole mass show little difference between SF- and AGN-dominated quasars in the low-mass regime (well below the median RQ quasar BH masses); on the other hand, the AGN-dominated quasar population tends to host more massive BHs than SF-dominated quasars in the high-mass regime, as shown by the difference in average BH masses for the SF- and AGN-dominated quasars. To quantify the difference between SF- and AGN-dominated quasar populations, we perform K-S tests between BH mass CDFs for SF- and AGN-dominated quasars in each grid cell, and the p-values are shown in Figure~\ref{fig:ecdf_mass_with_fit} for each grid cell. We are able to reject the null hypothesis that the two CDFs are identical (p-value<0.05) in most grid cells. \par

Our result is in agreement with the results of \citet{mclure_relationship_2004} and \citet{whittam_mightee_2022}. It also agrees with the result in Figure~\ref{fig:result_mass} where the quasars that host the most massive BHs are more likely to be radio-bright when compared to the rest of the population. Note that although $L>L_\textrm{pl}$ proves to be a good proxy for AGN-dominated sources in all grid cells, the exact shape of the cumulative distribution function can still be affected by minor features in the AGN tail of the radio flux density distribution that deviate from the single power-law assumption used in our model - for example, the SF- and AGN-dominated distributions arguably come closer together for quasar samples in more luminous $\mathcal{M}_i$ grid cells, where p-values rise slightly above 0.05. \par

Given the broadly consistent offset between SF- and AGN-dominated populations in the $\mathcal{M}_i-z$ space (as shown in Figure~\ref{fig:ecdf_mass_with_fit}), we here combine the $\mathcal{M}_i-z$ grid cells to get a more compact and detailed look at the results, while also investigating the impact of different RL / RQ classifications from the literature. \par

\begin{figure}     
    \includegraphics[width=\columnwidth]{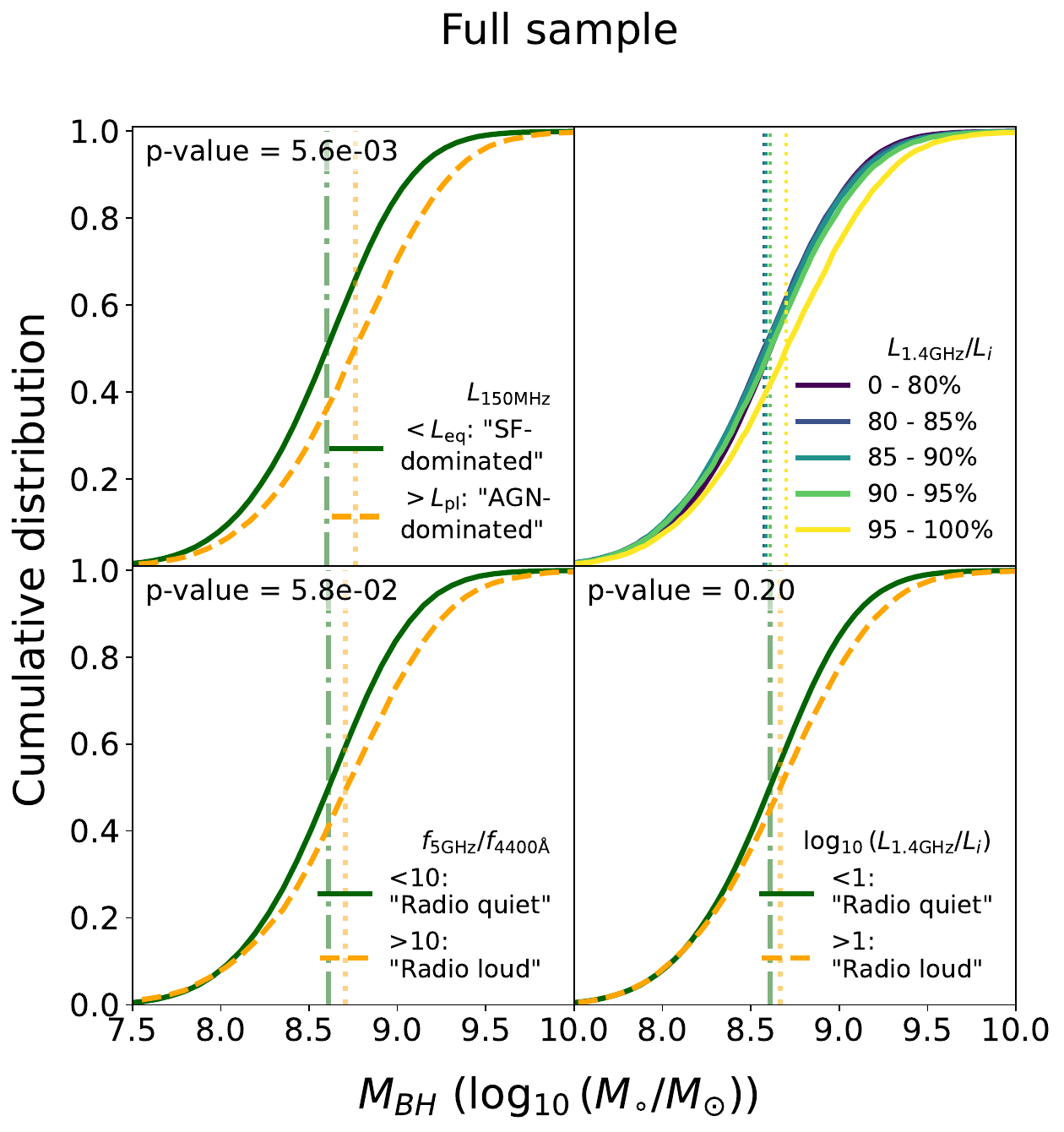}
    \caption{Difference in the BH mass CDF between RL/RQ (SF/AGN dominated) quasars under different definitions. We carefully examined the origin of quasar radio emission at different $M_\mathrm{BH}$ using our two-component model, and reproduce the difference in the cumulative BH mass distributions of SF- and AGN-dominated quasars (upper left). By binning quasars into different radio loudness percentiles (upper right), we show that only the top 5\% in quasar radio-loudness is responsible for the difference in the BH mass CDF. In the lower panels, quasars are classified into RL/RQ based on their flux density ratio \citep[top left; e.g.][]{mclure_relationship_2004}{}{} and luminosity ratio \citep[top right; e.g.][]{balokovic_disclosing_2012,arnaudova_exploring_2024}{}{} between radio and optical bands, with the thresholds listed in the corresponding panels. We quantify the difference of cumulative BH mass distributions between RL (AGN-dominated) and RQ (SF-dominated) quasars with p-values calculated from K-S tests performed on the two populations; smaller p-values indicate that the two distributions are less likely to be identical.}
    \label{fig:other}
\end{figure}

The black hole mass CDFs under our model-driven quasar classification are stacked across the $\mathcal{M}_i-z$ grid cells in Figure~\ref{fig:ecdf_mass_with_fit} and plotted in the top left panel of Figure~\ref{fig:other}. This is consistent with what we show in Figure~\ref{fig:ecdf_mass_with_fit}, with p-value=0.0054 rejecting the null hypothesis. In the bottom panels of Figure~\ref{fig:other}, we bin the same quasar sample set into RQ and RL quasars based on the different traditional radio loudness criteria mentioned above and plot the corresponding black hole mass CDFs for the RQ and RL population. We include the equivalent of Figure~\ref{fig:ecdf_mass_with_fit} for these classification criteria in Figure~\ref{fig:ecdf_mass} of Appendix~\ref{sec:appendix_ecdf} to give an overview of the evolution of their BH mass CDFs in the $\mathcal{M}_i-z$ space. \par

Specifically, in the bottom left panel of Figure~\ref{fig:other} we examine the radio loudness indicator used in \citet{mclure_relationship_2004}: $R_\mathrm{flux}=f_\textrm{5GHz}/f_\textrm{4400\AA}$, where the solid deep green line denotes the RQ quasars with $R_\mathrm{flux}<10$, and the dashed orange line shows the RL quasars with $R_\mathrm{flux}>10$. The mean BH mass for each population is shown as a vertical dotted line for RL quasars and a dash-dotted line for RQ quasars. Here, the 5 GHz radio flux density is converted from the LOFAR 150 MHz radio flux density by assuming a radio spectral slope of $\alpha=0.7$\footnote{We pick this value to keep in consistency with previous works including \citet{gurkan_lotsshetdex_2019}, \citet{smith_lofar_2021}, and \citet{arnaudova_exploring_2024}.}, and the $4400\textup{\AA}$ flux density is calculated from the fitted rest frame $3000\textup{\AA}$ continuum luminosity in \citet{wu_catalog_2022} by assuming an optical spectral index of $\alpha_\mathrm{opt}=0.5$. Despite the different radio-loudness indicators used, the trend remains consistent with Figure~\ref{fig:ecdf_mass_with_fit} where AGN-dominated quasars host more massive black holes in the high-mass regime, which is reflected by the difference in the mean BH mass of the RL and RQ samples. For comparison, the bottom right panel shows similar CDFs of RQ and RL quasar black hole masses under the classification criteria in \citet{balokovic_disclosing_2012} and \citet{arnaudova_exploring_2024} instead. Compared to the bottom left panel, the difference between the RL and RQ population is less significant (p-value=0.20), which is consistent with their conclusions. \par

\begin{table}
    \centering
    \caption{Number statistics of quasars being classified into RL / RQ under the luminosity definition ($R_\mathrm{lum}=\log_{10}\left({L_{1.4\textrm{GHz}}}/{L_i}\right)$) or flux definition ($R_\mathrm{flux}= {f_\textrm{5GHz}}/{f_\textrm{4400\AA}} $), and quasars being classified into SF-dominated (SF), radio intermediate (RI), and AGN-dominated (AGN) under the two-component model definition presented in this work. The bottom row shows the total number of quasars being classified into RL / RQ under either $R_\mathrm{lum}$ or $R_\mathrm{flux}$ definition, which are then used as the benchmark for calculating the percentage of RL /RQ quasars being classified into SF, RI or AGN in this work (shown in brackets below the numbers). The rightmost column shows the total numbers and percentages of the SF, RI, and AGN quasar population.}
    \label{tab:statistics}
    \begin{tabular}{llllll}
        \hline
        & \multicolumn{2}{c}{$R_\mathrm{lum}=\log_{10}\left(\frac{L_{1.4\textrm{GHz}}}{L_i}\right)$}&\multicolumn{2}{c}{$R_\mathrm{flux}= \frac{f_\textrm{5GHz}}{f_\textrm{4400\AA}} $}&\\
        & & & & &\\
        & RQ& RL& RQ& RL& Total\\
        \hline
        SF & 192717 & 1890 & 194566 & 41 & 194607\\
        & (94.2\%) & (10.4\%) & (91.7\%) & (0.4\%) &(87.4\%)\\
        RI & 10556& 6048& 14834& 1770& 16604\\
        & (5.2\%)& (33.3\%)& (7.1\%)& (16.5\%)& (7.4\%)\\
        AGN & 1328& 10243& 2672& 8899& 11571\\
        & (0.6\%)& (56.3\%)& (1.2\%)& (83.1\%)& (5.2\%)\\
        \hline
        Total & 204601 & 18181 & 212072 & 10710 & 222782\\
        & (100\%) & (100\%)  & (100\%) & (100\%) & (100\%)\\
    \end{tabular}
\end{table}

Going into more detail, Table~\ref{tab:statistics} compares the number statistics of quasars classified into different populations according to different criteria. Each row shows one of the three quasar populations under our model-driven definition: SF-dominated (SF), radio intermediate (RI), and AGN-dominated (AGN); the columns outline the classification under the luminosity ($R_\mathrm{lum}=\log_{10}\left({L_{1.4\textrm{GHz}}}/{L_i}\right)$) and flux ($R_\mathrm{flux}=  {f_\textrm{5GHz}}/{f_\textrm{4400\AA}} $) definition, where $R_\mathrm{lum}<1$ ($R_\mathrm{flux}<10$) quasars are defined as RQ and $R_\mathrm{lum}>1$ ($R_\mathrm{flux}>10$) quasars are defined as RL. The luminosity definition provides a clean selection of SF-dominated quasars in their RQ population, with 94.2\% of the RQ quasars dominated by SF; only 0.6\% of which are AGN-dominated. However, the single-value threshold introduces a fair amount of SF-dominated and radio intermediate quasars into the RL population, with only 56.3\% of the RL quasars are being securely AGN-dominated. This contaminated selection significantly reduces the BH mass difference between the SF- and AGN-dominated quasars when they compared their RL samples with RQ samples. The flux definition, on the other hand, provides a cleaner selection in their RL population, with 83.1\% of them being AGN-dominated. However, their RL quasar selection also misses out 2672 AGN-dominated quasars, which accounts for 23.1\% of the entire AGN-dominated quasar population - twice the number of the 1328 dropouts under the luminosity definition. Despite the incomplete selection of AGN-dominated quasars, RL and RQ populations under the flux definition still reflect the difference in their BH mass distribution, since the number of contamination sources introduced into the RL sample is 26.8\% less than the luminosity definition. \par

To better understand these differences, in the upper right panel of Figure ~\ref{fig:other} we divide the quasars by their percentiles in the radio loudness distribution (here defined as $R_\mathrm{lum}=L_\textrm{1.4GHz}/L_i$), and the quasars within the top 20\% of the radio loudness distribution are further binned into $5\%$ wide percentile bins. This is motivated by the definition of RL/RQ AGNs in \citet{whittam_mightee_2022}, where they argued that AGNs at 1-$\sigma$ beyond the infrared-radio correlation (IRRC) - which corresponds to the 70th percentile in the $L_\textrm{radio}/L_\textrm{bol}$ distribution - are probably not powerful enough to be classified as RL, and they adopted the 2-sigma radio excess - which corresponds to the 95th percentile in the distribution - as the threshold for RL AGNs. \par

From Figure~\ref{fig:other}, we can see that the difference in BH mass CDF occurs only in $5\%$ of the quasar population in this study; this fraction is consistent with the fraction of AGN-dominated quasars in the entire sample (5.6\%; see Table~\ref{tab:statistics}). This confirms that the differences in BH mass only occur in the AGN tail of the radio flux density distribution, not the SF population. Furthermore, as noted earlier, this difference in BH mass CDF only becomes more pronounced for BH with masses larger than $\log(M/M_\odot)=8.5$, which is in line with the trend presented in Figure~\ref{fig:result_mass} where the quasars with the most massive BHs tend to be more radio loud. These results show that AGN-dominated quasars have a $M_\mathrm{BH}$ distribution that differs from their SF-dominated counterparts, suggesting a connection between $M_\mathrm{BH}$ and jet production. \par

\begin{figure}     
    \includegraphics[width=\columnwidth]{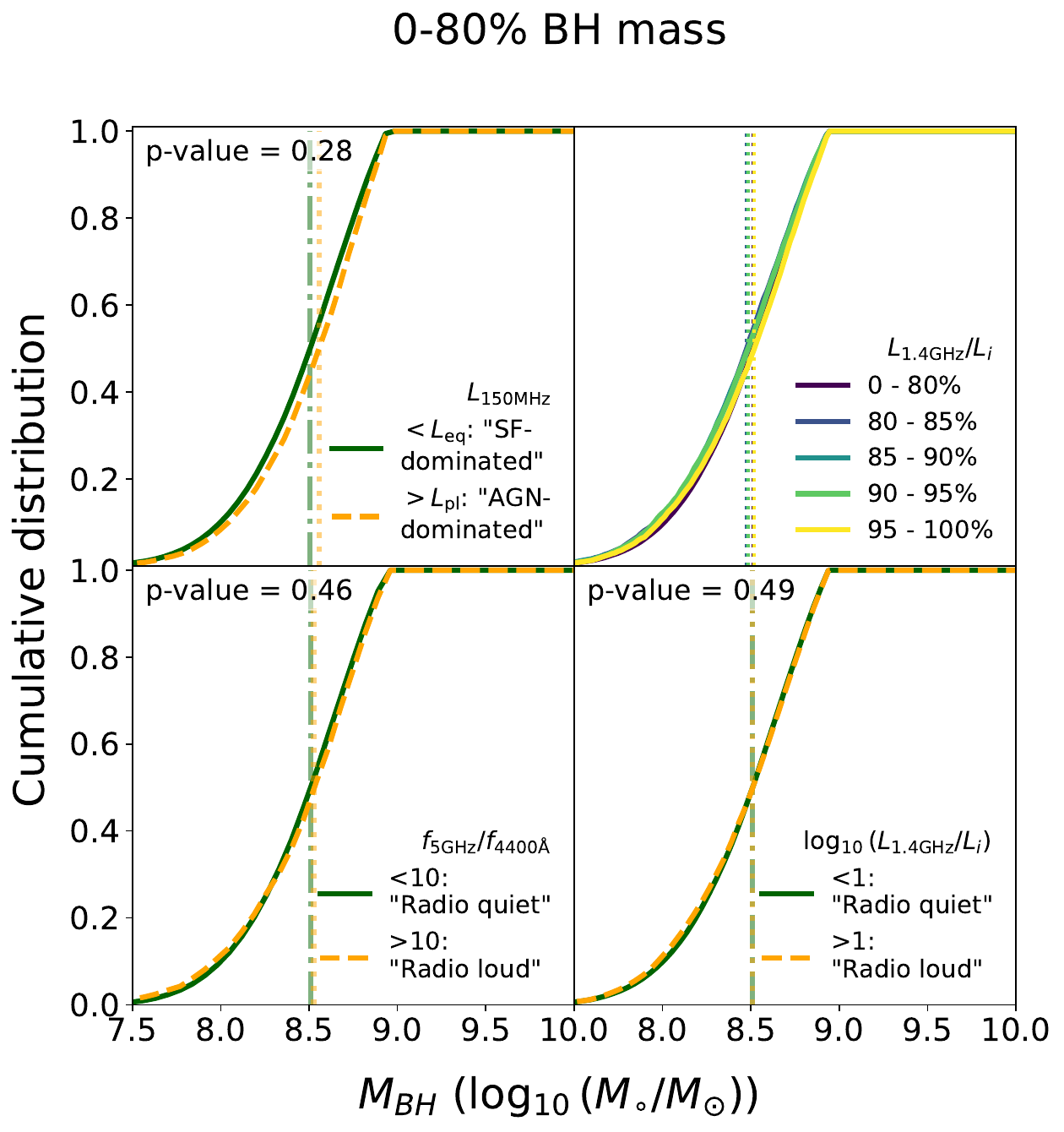}
    \caption{The black hole mass cumulative distribution function after removing the top 20\% in the black hole mass distribution, separated into RL (AGN-dominated) and RQ (SF-dominated) quasars based on various definitions of radio loudness. In contrast to Figure~\ref{fig:other}, there is no difference between RL and RQ population under all radio loudness definitions (p-value>0.05), which confirms that the excess in quasar radio loudness only comes from the most massive (top 20\%) quasars. }
    \label{fig:other_masscut}
\end{figure}

Finally, we check the consistency between this analysis and the earlier analysis from our Bayesian model fitting. In Figure~\ref{fig:result_mass} we had shown that among the AGN-dominated subset, it was only the quasars within the top 20\% of the BH mass distribution in each grid cell that contributed to the excess radio emission. Therefore, in Figure~\ref{fig:other_masscut}, we remove all quasars in the top 20\% of the BH mass distribution in each grid cell, and again plot the BH mass CDFs from Figure~\ref{fig:other}. By applying the mass cut, the black hole mass distribution becomes consistent between RL and RQ quasars under all classification methods, with p-values rising above 0.2 for all criteria. This is in full consistency with our results from Figure~\ref{fig:result_mass}. We therefore further secured the origin of the excessive radio loudness associated with black hole mass down to the most massive (top 20\% in the black hole mass distribution) and most radio loud (top 5\% in the radio loudness distribution, dominated by the AGN component) quasars alone. \par

Combining the analysis from Section~\ref{sec:result} and \ref{sec:analysis}, we can conclude that under most scenarios there is no fundamental connection between quasar radio emission and BH mass since the radio emission from the host galaxy does not depend on $M_\mathrm{BH}$, and radio emission from the AGN remains the same across most BH mass percentiles; however, for the `genuine' RL quasars where powerful AGN jets dominate the quasar radio emission, there is a boost in the level of AGN jet activity in quasars above a certain BH mass threshold. Since all of the panels in Figure~\ref{fig:other} trace the BH mass distribution of the same quasar population, the disagreement between previous results originates solely from definitions of quasar radio loudness that fail to reflect the physical processes behind the scene. More specifically, the reason that \citet{gurkan_lotsshetdex_2019} and \citet{arnaudova_exploring_2024} did not identify a strong dependence on $M_\textrm{BH}$ between the RQ and RL populations is that the assumed RL threshold classified $\sim30\%$ radio intermediate sources into radio loud when their radio emission is not necessarily dominated by the AGN component, and that the $R_\mathrm{lum}-M_\mathrm{BH}$ distribution is dominated by the SF-dominated quasars; as a result, the BH mass distribution of the RL samples in \citet{arnaudova_exploring_2024} shows a higher level of similarity to their RQ samples, and \citet{gurkan_lotsshetdex_2019} failed to find any correlation between $R_\mathrm{lum}$ and $M_\mathrm{BH}$. Note that \citet{arnaudova_exploring_2024} reports a slightly higher $M_\textrm{BH}$ for RL quasars when using \ion{H}{$\beta$} and [\ion{O}{iii}] line measurements, which is also in line with our result in Figure~\ref{fig:other}. On the other hand, the RL quasar population defined in \citet{mclure_relationship_2004}, although incomplete, was not significantly contaminated by non-AGN-dominated quasars, resulting in a difference in the BH mass distribution between RL and RQ quasars. Our new understanding of traditional radio loudness definitions helps unify these different observational results into a coherent framework.\par

\section{Testing for AGN outflow signals}
\label{sec:test}

Outflow structures in quasar emission lines can lead to systematically higher measured FWHMs (and hence BH masses) within the published SDSS catalogues, as shown in \citet{coatman_correcting_2017} for \ion{C}{iv} BH mass measurements. Although \ion{C}{iv} BH masses are already excluded from our sample, quasars with the most extreme outflow activity could still be misclassified into more massive BH mass bins (i.e. with BH masses being overestimated); in turn, this would bring a radio excess into the quasar population hosting most massive BHs, since outflow activities are associated with higher radio detection fraction of quasars \citep[e.g.][]{richards_probing_2021,petley_how_2024,escott_unveiling_2025}{}{}.

To examine the potential observational bias of the BH mass in our quasar sample and to quantify the impact of AGN outflow activities on our conclusion, we perform an analysis of the stacked properties of quasars in our sample set. We start by separating our quasars into four quadrants based on their BH masses in the SDSS catalogue and the dominant source of their radio emission: \par

\begin{enumerate}
    \item Quadrant 1: Quasars hosting moderate mass BHs (0 - 80\% in the BH mass distribution of quasars within the corresponding $M_i-z$ grid cell) with radio emission dominated by SF activity;
    \item Quadrant 2: Quasars hosting massive BHs (top 20\% in the BH mass distribution of quasars within the corresponding $M_i-z$ grid cell) with radio emission dominated by SF activity;
    \item Quadrant 3: Quasars hosting moderate mass BHs with radio emission dominated by AGN activity;
    \item Quadrant 4: Quasars hosting massive BHs with radio emission dominated by AGN activity.
\end{enumerate}

In total, 157,229 quasars are classified into quadrant 1, 37,378 quasars are classified into quadrant 2, 8,156 quasars are classified into quadrant 3, and 3,415 quasars are classified into quadrant 4. The rest of quasars either are defined as radio intermediates or have spurious BH mass measurements ($M_\mathrm{BH}\leq0$) in the SDSS DR16Q catalogue and are thus not included in the classification. \par

To robustly compare the difference in physical properties between different quadrants while ensuring that the differences are not driven by underlying correlations with $M_i$ or $z$, we match the quasars of different quadrants in the $M_i-z$ space. We start by defining the distance in the $M_i-z$ space as $d=\sqrt{\Delta M_i^2+(\epsilon\Delta z)^2}$, where the coefficient $\epsilon$ is defined as the length ratio of each $M_i-z$ grid cell, $\epsilon=1/0.4$, to compensate for the relative difference in the scale of $M_i$ and $z$ in the Euclidean space. Next, we define the `closest neighbour' pair as two quasars having the smallest distance in the $M_i-z$ space. We then match every quasar in quadrant 4 with their closest neighbour in quadrants 1, 2, and 3. The matched quasars in quadrants 1, 2, and 3 therefore share the same redshift and bolometric luminosity properties as quasars of quadrant 4\footnote{In the remaining part of this section, we will use quadrant 1 to 4 quasars to refer to the \emph{matched} quasar population within each quadrant.}. We then use the matched quasars in quadrants 1, 2, 3, and 4 for our analysis below.\par

First, we focus on the potential bias in the BH mass measurements by investigating the difference in the stacked spectra of the \ion{Mg}{ii} emission line, which is used to obtain single-epoch BH mass estimations in our sample. \par

As briefly discussed above, strong AGN outflows will be visible from the shape of the emission lines as an extended wing on the blue end. Such line asymmetry will lead to an overestimation in the single-epoch emission line FWHM and hence to an overestimation in the BH mass. Although the individual quasar spectra from SDSS DR16Q are often noisy, we can take advantage of the large size of our quasar sample to test for such a systematic effect by stacking the spectra of quasars with similar redshifts and luminosities. We select a narrow ($\Delta \log M_\mathrm{BH} = 0.2~\textrm{dex}$) range of BH mass at the high $M_\mathrm{BH}$ end, and stack the quadrant 2 and 4 quasars in that $M_\mathrm{BH}$ range for different $M_i-z$ grid cells. We then compare the stacked emission line profiles to check for any asymmetry caused by SF- or AGN-dominance in quasars. \par

\begin{figure}     
    \includegraphics[width=\columnwidth]{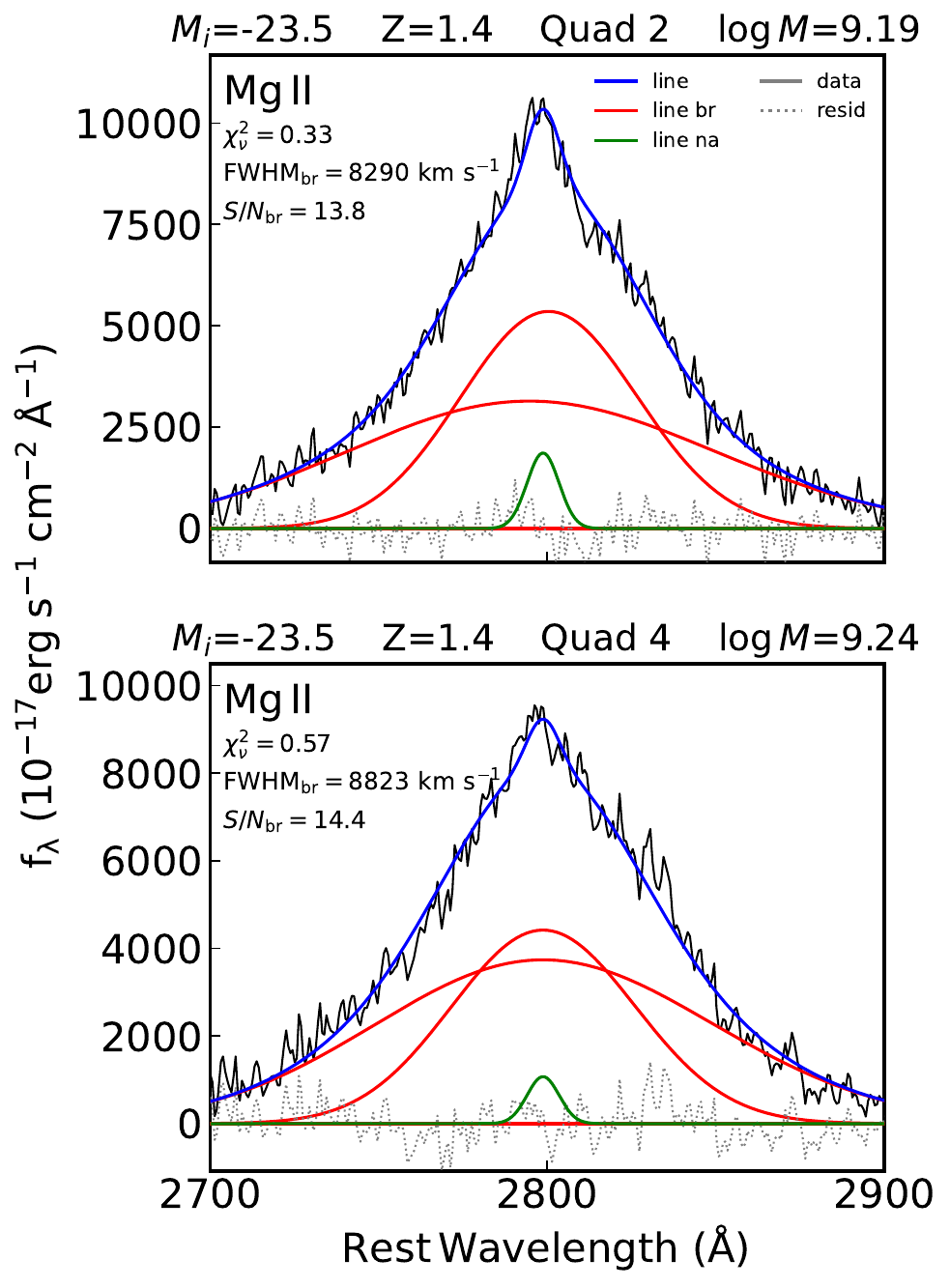}
    \caption{We show one representative $\mathcal{M}_i-z$ bin ($-24<\mathcal{M}_i<-23$, $1.2<z<1.6$) in which we stack the SDSS spectra of quasars that are matched in $\mathcal{M}_i-z$ space and have catalogue BH masses of $9.1<\log (M_\mathrm{BH}/M_{\odot})<9.3$ (within the highest 20\% BH mass bin), while being dominated by SF (quadrant 2, upper panel) and AGN (quadrant 4, lower panel) components, respectively. The stacked spectra (grey solid line) appear very similar. Both are then fitted with PyQSOFIT, with the best fits of broad emission line components and narrow line component plotted in red and green solid lines respectively; the residual of the fit is plotted as grey dotted line. The blue solid line shows the best fit to the \ion{Mg}{ii} line, combining broad and narrow components. The BH mass values shown on top of each panel is measured from the stacked spectra, calculated with the broad line FWHM best fit (shown in the legend) using the recipe in \citet{vestergaard_mass_2009}; the BH masses from stack broadly recovers the median input value.}
    \label{fig:mgii_stack}
\end{figure}

We find that the \ion{Mg}{ii} emission line profiles are similar across all $M_i-z$ grids, with no evidence of systematically different outflow components in AGN-dominated (quadrant 4) quasars compared to SF-dominated (quadrant 2) quasars, indicating that outflow features in radio-bright quasars do not have a systematic effect on the BH mass measurements in our sample. As an example, within a representative $\mathcal{M}_i-z$ grid cell, we select quadrants 2 and 4 quasars with a catalogue BH mass of $9.1<\log (M_\mathrm{BH}/M_{\odot}) <9.3$, and we show the \ion{Mg}{ii} components of the stacked SDSS spectra of these quasars in Figure~\ref{fig:mgii_stack}. Here, we adopt the \ion{Mg}{ii} BH masses from \citet{wu_catalog_2022} when selecting quasars in the stack. The average catalogue values of the BH mass are $\log (M_\mathrm{BH}/M_{\odot})=9.19$ for both stacks. To test the recovery of $M_\mathrm{BH}$, we fit the entire spectrum with the PyQSOFIT code \citep{shen_sloan_2019} while adopting the \ion{Mg}{ii} BH mass recipe from \citet{vestergaard_mass_2009}:

\begin{multline}
    \log (M_\mathrm{BH}/M_\odot)=6.86+2~\log\left(\mathrm{\frac{FWHM(\ion{Mg}{ii})}{1000~\mathrm{km\cdot s}^{-1}}}\right)\\
    +0.5~\log\left(\frac{L_{3000}}{10^{44}~\mathrm{erg\cdot s}^{-1}}\right),
\end{multline}

\noindent where $L_{3000}$ is the monochromatic luminosity $\lambda L_\lambda$ at 3000\AA. When computing the fitted BH masses, we use the best fit values from PyQSOFIT for $\mathrm{FWHM(\ion{Mg}{ii})}$, and we average the $L_{3000}$ values obtained from \citet{wu_catalog_2022} for quasars in the stack as input for $L_{3000}$. We find that recovered BH masses are very close to the median input mass. For the specific example shown in Figure~\ref{fig:mgii_stack}, the stacked \ion{Mg}{II} line for quadrant 4 quasars has a higher FWHM than that measured with quadrant 2 quasars, leading to a $0.05~\mathrm{dex}$ difference between the \ion{Mg}{ii} BH mass calculated from the stack ($\log M_\mathrm{BH}/M_\odot=9.24$) and the average \ion{Mg}{ii} BH mass from catalogue. This suggests that the BH masses for quasars in quadrant 4 might actually be underestimated in this case (the opposite of the potential bias we were concerned about). \par

Next, we examine the interplay between the level of outflow activity, BH mass, and the dominant source in radio emission using the properties of \ion{C}{iv} emission line as a tracer for AGN outflow. Recent works \citep[e.g.][]{richards_unification_2011,richards_probing_2021,rankine_bal_2020,temple_testing_2023} have explored the diagnostic power of 2D \ion{C}{iv} equivalent width (EW) and blueshift space through the definition of `\ion{C}{iv} distance'\footnote{We refer the readers to \citet{richards_probing_2021} for the exact definition and calibration of \ion{C}{iv} distance used in this work.}. In \citet{richards_unification_2011}, the 2D parameter space of \ion{C}{iv} EW and blueshift is proposed to have a connection with quasar accretion modes, where disc-dominated accretions have higher \ion{C}{iv} EW and low blueshifts, whereas wind-dominated accretions tend to reside in the parameter space with low \ion{C}{iv} EW and high blueshifts. \citet{richards_probing_2021} further quantified this difference in \ion{C}{iv} space by converting the 2D distribution into a single value of \ion{C}{iv} distance. When calibrated against the SDSS-RM quasar samples, the \ion{C}{iv} distance itself can act as a proxy for `disc' versus `wind' activity \citep{richards_unification_2011,rankine_bal_2020}, with lower \ion{C}{iv} distance being associated with `disc' model and higher \ion{C}{iv} distance to `wind' model. Through extensive modelling of quasar SEDs, \citet{temple_testing_2023} quantified the differences in \ion{C}{iv} and \ion{He}{ii} emission line properties and the underlying theoretical accretion model within the Eddington ratio ($L/L_\mathrm{Edd}$) and $M_\mathrm{BH}$ plane. Quasars with the highest \ion{C}{iv} EW are found in the parameter space with a high Eddington ratio ($L/L_\mathrm{Edd}\gtrsim0.2$) and a relatively low BH mass ($\log (M_\mathrm{BH}/M_{\odot})<9.0$), or with a low Eddington ratio and a high BH mass. Quasars with high \ion{C}{iv} blueshifts are observed in a different part of the parameter space with a high Eddington ratio and a high BH mass. \citet{temple_testing_2023} also suggested a likely discrepancy between the existing simple SED models and the observation when the Eddington ratio drops below $L/L_\mathrm{Edd}\approx0.2$, which may be the result of a transition phase in the accretion geometry \citep{giustini_global_2019}. \par 

\begin{figure}     
    \includegraphics[width=\columnwidth]{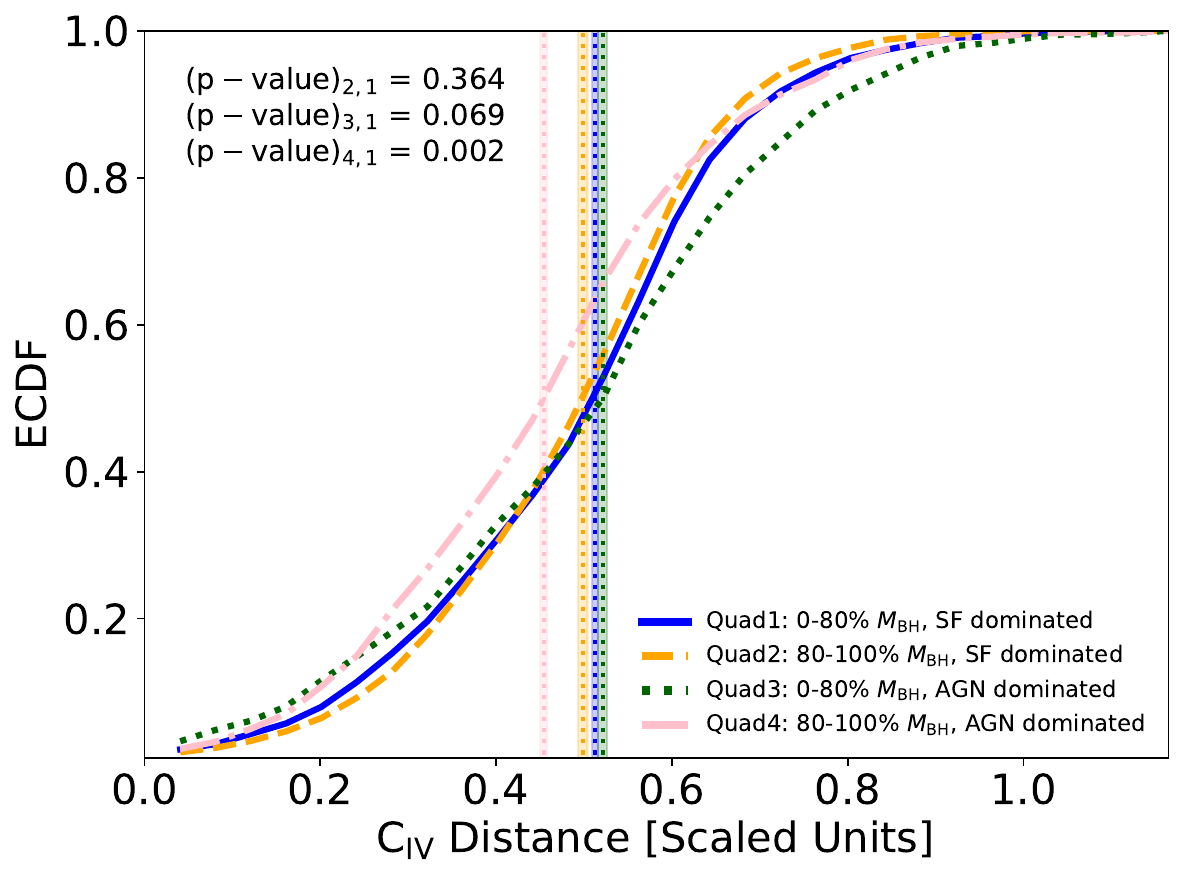}
    \caption{The cumulative distribution function of quasar \ion{C}{iv} distances in quadrants 1 to 4, calculated from the line of best-fit in \citet{richards_probing_2021}. Note that this figure does not include quasars with $\log \ion{C}{iv}~\mathrm{EW}>2.5$ as they fall out of the domain of definition in \citet{richards_probing_2021}. Quadrants 1 to 4 are marked with blue solid line, orange dashed line, green dotted line and pink dashed-dotted line, respectively. The mean values of each distribution are marked as solid dotted lines, and the uncertainty is determined from the 16\% and 84\% confidence intervals when bootstrap sampling the distributions. We have performed K-S tests between the CIV distance distribution from quadrant 1 and the rest of quadrants, and the corresponding p-values are shown in the upper left corner of the plot. While quadrants 2 and 3 show a higher similarity with quadrant 1, the p-values between quadrant 1 and 4 falls below 0.05, hence we reject the null hypothesis that the \ion{C}{iv} distance distributions from these two quadrants are identical.}
    \label{fig:civ_distance}
\end{figure}

To characterise the distribution of quasars from the four quadrants in the \ion{C}{iv} space, we use the recipe from \citet{richards_probing_2021} to compute \ion{C}{iv} distances for every quasar in our matched sample where \ion{C}{iv} emission lines are visible from SDSS spectra. Figure~\ref{fig:civ_distance} shows the cumulative distribution function of \ion{C}{iv} distances for quasars within each quadrant, where quadrants 1 to 4 are colour coded with blue solid line, orange dashed line, green dotted line and pink dashed-dotted line, respectively. The vertical lines show the average values within each quadrant, and the uncertainty is determined by bootstrap resampling of the sources. The difference between the distributions is quantified by the p-values of the K-S tests performed between quadrant 1 and the other quadrants. \par

For the SF-dominated quasars in quadrants 1 and 2, their distribution in the \ion{C}{iv} space shows minimal difference (p-value>0.05) since their \ion{C}{iv} distance distributions are almost identical, suggesting that a common accretion mode is shared between all SF-dominated quasars. AGN-dominated quasars with moderate BH mass are more likely to reside at the high \ion{C}{iv} distance end (\ion{C}{iv} distance $>0.5$) than SF-dominated quasars. The most significant difference is between the AGN-dominated quasars with high BH mass and the rest of the populations: quadrant 4 quasars tend to reside in the lower \ion{C}{iv} distance parameter space, which is reflected by both a lower average \ion{C}{iv} distance than the rest and a low p-value (0.002) between the \ion{C}{iv} distance distribution of quadrant 4 and quadrant 1 quasars, which in principle rejects the null hypothesis. However, note that our analysis does not take into account the significant uncertainties in individual measurements. The difference in quadrant 4 quasars suggests that these systems are less outflow dominated than the other quasar populations and that there is no evidence for an enhancement in AGN outflow in the AGN-dominated high-BH mass sample. This agrees with an increased radio detection fraction at the low \ion{C}{iv} distance end reported in \citet{richards_probing_2021}. On the other hand, \citet{richards_probing_2021} also discovered an increase of radio detection fraction at the high \ion{C}{iv} distance end; since the AGN-dominated sources under our classification are by definition jet-dominated, such an increase can instead be attributed to the dominance of AGN wind and/or outflow in the radio emission of radio intermediate quasars.

\begin{figure}     
    \includegraphics[width=\columnwidth]{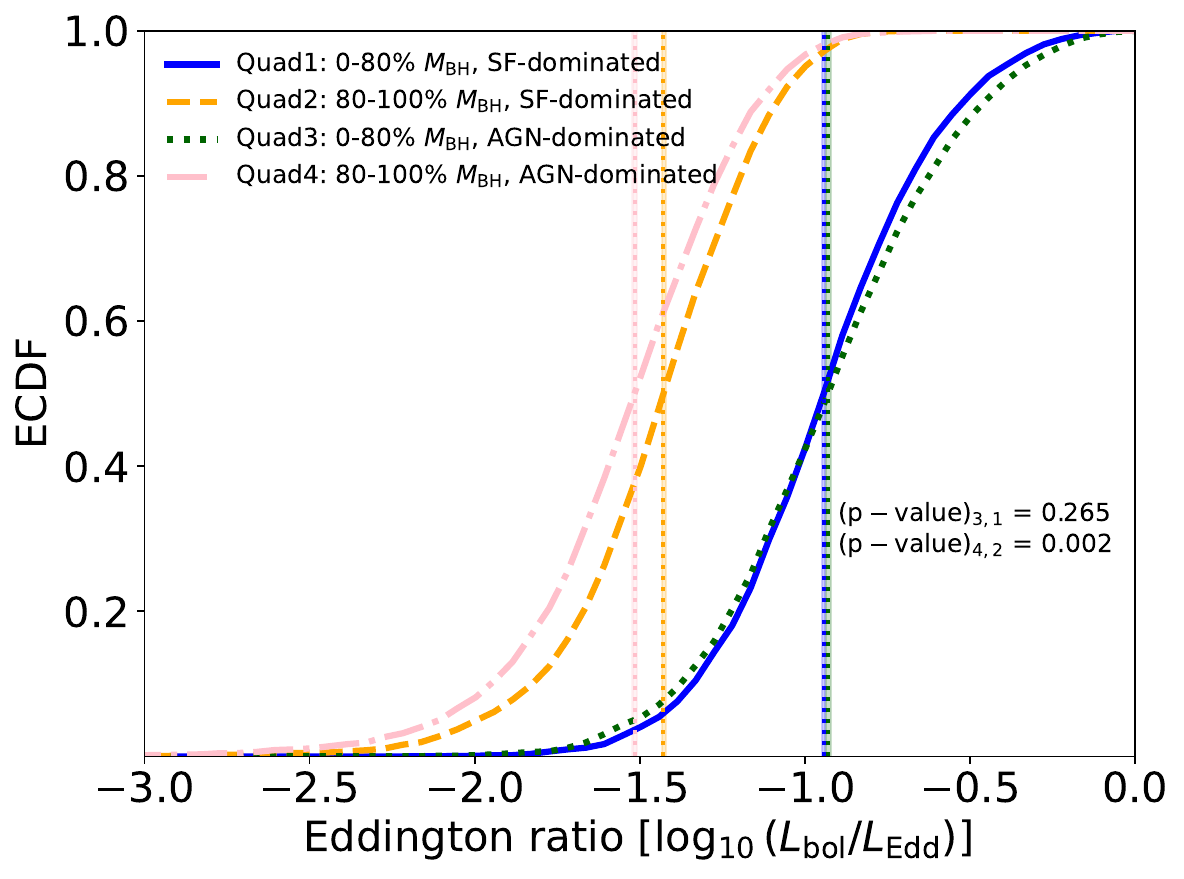}
    \caption{The cumulative distribution of Eddington ratio in quadrants 1 to 4 quasar populations, plotted in blue solid line, orange dashed line, green dotted line and pink dashed-dotted line, respectively. The vertical lines mark the mean value of Eddington ratio in each population. The p-values at the lower right corner is calculated from the K-S tests performed between quadrants 1 and 3, and quadrants 2 and 4, respectively. The Eddington ratio distribution is identical between quadrants 1 and 3; however we reject the null hypothesis for quadrants 2 and 4.}
    \label{fig:edd_ecdf}
\end{figure}

We also investigated the difference in the BH accretion efficiency traced by the Eddington ratio ($L/L_\mathrm{Edd}$). Note that our samples are matched in quasar luminosity and that the Eddington ratio scales as $L/M_\mathrm{BH}$, therefore this effectively traces the difference in $M_\mathrm{BH}$ between different quadrants. Hence, quasars with moderate BH mass (quadrants 1 and 3) inherently have a higher $L/L_\mathrm{Edd}$ than quasars with high BH mass (quadrants 2 and 4), due to selection biases. \par

Figure~\ref{fig:edd_ecdf} plots the distribution of the Eddington ratio for quasars matched in the four quadrants, with the same colour code as in Figure~\ref{fig:civ_distance}. For quasars with moderate BH masses, the SF-dominated and the AGN-dominated populations show no difference in their Eddington ratio distributions; this is expected since these quasars are matched in luminosities and redshifts while occupying the same BH mass percentiles. However, for quasars with massive BHs, the Eddington ratio distribution of AGN-dominated quasars is different from SF-dominated quasars (p-value = 0.002), and AGN-dominated quasars with high BH masses tend to have lower Eddington ratios in general. This agrees with the findings in Section~\ref{sec:analysis} where AGN-dominated quasars tend to be more massive than SF-dominated quasars (hence the lower Eddington ratio in matched samples), and such difference only occurs in quasars hosting the 20\% most massive BHs. \par

Combining the results in Figure~\ref{fig:civ_distance} and \ref{fig:edd_ecdf}, we can map the quadrant 4 quasars to the quasar population highlighted in \citet{temple_testing_2023} with low \ion{C}{iv} distances, low Eddington ratio and high BH masses. This suggests that alternative accretion physics in the broad line region \citep[e.g.][]{mitchell_soux_2023,temple_testing_2023} may play a role in the uniqueness of this AGN-dominated massive quasar population.

\section{Discussion}
\label{sec:discussion}

We have investigated the relationship between BH mass and quasar radio loudness from two different perspectives: in Section~\ref{sec:result} we binned the quasars by their BH mass and found that quasars with the most massive BHs have an excess in radio emission, while in Section~\ref{sec:analysis} we show that quasars with extreme radio excess are also more likely to host more massive BHs. Combining the analysis from these two sections, we have singled out the quasar population that is the top 5\% in radio loudness and top 20\% in BH mass; only quasars of these highest black hole masses show a different distribution of radio flux densities, and this distribution only differs within the top 5\% of the radio luminosities, corresponding to the radio luminosities dominated by the AGN / jet component. This suggests a difference in the nature of these quasars compared to the rest of the population. In Section~\ref{sec:test} we have gone on to show that this is not driven by a bias in $M_\mathrm{BH}$ measurements; yet these quasars do show different \ion{C}{iv} line equivalent widths, blueshifts and Eddington-scaled accretion rates than the rest of the quasar sample. \par

Although the data analysed in this paper do not allow us to draw a robust conclusion on the origin of this difference in radio flux densities, in this section we consider a variety of different mechanisms or effects that might be driving this difference: \par

\begin{enumerate}
    \item BH accretion mode. Low accretion rate BHs are thought to host hot accretion flows, e.g. the advection-dominated accretion flow \citep[ADAF;][]{narayan_advection-dominated_1994,narayan_advection-dominated_1995}, while high accretion objects host cold accretion flows generated from thin disk accretion. AGNs are then classified into jet-mode and quasar-mode (radiative-mode) based on their accretion mode \citep[e.g.][]{heckman_coevolution_2014}. Previous studies have constructed a fundamental plane of jet-mode AGNs using BH mass, core jet radio luminosity and disk X-ray luminosity \citep[e.g.][]{merloni_fundamental_2003}{}{}, which then links the radio jet production to the accretion rate and black hole mass. However for high accretion rate objects including radio quasars, such a fundamental plane is not well established \citep[see e.g.][]{merloni_synthesis_2008,plotkin_using_2012,inoue_disk-jet_2017}{}{}, implying multiple BH accretion modes may be at play. In \citet{ichikawa_quenching_2017}, the authors suggested that for radio quasars, the difference between the RL fraction under different $M_\textrm{BH}$ may be linked to the alternative jet production mechanism of AGNs, where a large enough BH mass is required for a powerful ($L_\mathrm{mech}>L_\mathrm{bol}$) radio jet to form and maintain itself. Under this scenario,  the gas flow at the core region of the SMBH converts into an ADAF from the original accretion disk. This suggests a transitional phase where a jet with enough mechanical power can inject itself into the ambient gas background and form shock regions where the radio emission is generated by non-thermal electrons, while the exterior of the accretion disk remains stable, or the AGN accretion disk becomes fragmented due to gravitational instability. As the formation of ADAF requires an Eddington fraction below $\sim0.1$, for sources with a given bolometric luminosity this leads to a high BH mass. Studies have found a strong BH mass dependence in the jet-mode AGN activity \citep[see][and references therein]{kondapally_cosmic_2022}{}{}. Meanwhile, \citet{sikora_magnetic_2013} also suggested a mixture of accretion modes in the triggering of radio loud quasars, but through an alternative path where an episode of hot accretion is followed by a massive, cold accretion event. As a result, the large magnetic flux is trapped within the inner radius of the newly formed cold accretion disk and forms a magnetically choked accretion flow \citep[MCAF;][]{narayan_magnetically_2003,reynolds_trapping_2006,mckinney_general_2012}{}{}. This path is most likely associated with major merger activities involving a giant elliptical galaxy and a disk galaxy.
    
    \item BH spin. It is widely believed that quasar radio jets are generated from a combination of magnetic field and rotational energy \citep[e.g.][the `BZ' mechanism]{blandford_electromagnetic_1977}; under this scenario the jet power is correlated with BH spin. Results from general relativistic magnetohydrodynamic simulations \citep[e.g.][]{hawley_magnetically_2006,martinez-sansigre_evidence_2011}{}{} helped constrain the BH spin influence on the radio jet power from the theoretical side. Meanwhile, on the observational side, \citet{martinez-sansigre_observational_2011} parameterised the spin distribution of black holes with $\log_{10}(M_\textrm{BH}/M_\odot)>8$ from radio observations, and found a small population of high excitation galaxies with both maximal spin (inferred from parameters associated with BH spin; indicating high radio loudness) and high accretion efficiency. \citet{schulze_evidence_2017} used the [\ion{O}{iii}] luminosity as an indirect tracer of BH spin, and suggested that RL quasars have systematically higher radiative efficiencies than RQ quasars, therefore higher BH spin. However, in these works the sample selection is biased by the detection limit of FIRST survey, and their estimation of black hole spin is limited by model uncertainties. Previous studies have associated high BH spin objects with major merger events \citep[e.g.][]{wilson_difference_1995,chiaberge_origin_2011,chiaberge_radio_2015}, implying a higher BH mass when compared to low spin BHs. Although whether BH spin governs the quasar radio loudness is still under debate \citep[see e.g.][]{sikora_magnetic_2013,inoue_disk-jet_2017,macfarlane_radio_2021}{}{}, an alternative scenario could be where the power from magnetically-driven outflow is distributed differently between the jet and disc component due to a combined effect of both BH spin and magnetic field configuration, as shown by \citet{mehdipour_relation_2019} through performing high-resolution X-ray spectroscopy on radio-loud AGNs, albeit with a small sample size. They have found a inverse correlation between the AGN wind and jet components (radio loudness), and also predicted disc truncation as a likely consequence of this scenario.
    
    \item Environmental factors. Studies including \citet{mandelbaum_halo_2009}, \citet{ross_clustering_2009} and \citet{shen_quasar_2009} show that radio loud quasars tend to reside in more massive haloes and denser environments, while the link between quasar environment, BH mass (stellar mass) and jet power remains unclear. Further evidence for this was provided by \citet{retana-montenegro_probing_2017} who used sources from SDSS and FIRST surveys and found that RL quasars are more strongly clustered than RQ quasars, while quasars powered by the most massive BHs also cluster more strongly than quasars with less massive BHs. It has been argued that denser environments can lead to a boosting of the radio luminosity (for fixed jet power) due to the confinement of radio lobes preventing energy losses of the relativistic electrons in radio lobes through adiabatic expansion \citep[e.g.][]{barthel_anomalous_1996}{}{}; in addition \citet{hatch_why_2014} argues that the dense environment fosters the formation of radio jets. However the conclusions of these studies could be affected by the different selection bias and definitions of radio loudness in AGN samples used by different studies (as highlighted in our analysis above) and the contamination from star-forming galaxies \citep[see e.g. discussion in][]{magliocchetti_hosts_2022}.
\end{enumerate}

The nature of this extreme quasar population remains an interest for future studies, since the data required to distinguish between potential scenarios that explain such differences are not yet available. Robust measurements of the quasar emission line and continuum through next-generation optical spectroscopic surveys such as WHT Enhanced Area Velocity Explorer Surveys \citep[WEAVE;][]{jin_wide-field_2024}{}{} - specifically the WEAVE-LOFAR survey \citep[][]{smith_weave-lofar_2016}{}{} - and the Black Hole Mapper programme in SDSS-V will provide the necessary information to identify observational biases in the current quasar sample and provide extended BH mass calibrations. Together with accurate photometric measurements of host galaxy morphologies, stellar mass, and environmental factors from \textit{Euclid} \citep[][]{euclid_collaboration_euclid_2024}, we can gather critical evidence to distinguish between these scenarios and determine the physical mechanism behind the radio excess in the most massive and radio-loud quasars.

\section{Summary}
\label{sec:conclusion}

In this work, we extend the fully Bayesian two-component model presented in \citetalias{yue_novel_2024} to study the evolution of host galaxy star formation and AGN jet activities with BH mass. We model the quasar radio emission by assuming that every quasar hosts both SF activity through their host galaxies (characterised by mean SFR $\Psi$) and jet activity through central BHs (characterised by AGN jet normalisation $f$). We bin the quasar samples into grid cells based on their location in the parameter space of $\mathcal{M}_i$, $z$ and BH mass, before fitting the radio flux density distribution within each grid cell against our two-component model to get the best fits for host galaxy SFR and AGN jet normalisation (relative AGN activity level compared to host galaxy SF) in our quasar sample. Both host galaxy and jet activities show little difference in most of the BH mass bins; however, we detect an excess in the jet activity level for quasars hosting the most massive (top 20\% in $M_\textrm{BH}$) black holes, where the normalisation of the power-law jet luminosity function ($f$) is increased by a factor of $\sim2.5$, suggesting these quasars are 2.5 times more likely to host powerful jets when compared to the rest. 

In order to build a coherent picture on the relation between quasar radio emission and BH mass that explains the differences in the previous studies on the BH mass dependence of radio loudness, we present a series of new definitions on the radio quasar population - AGN-dominated ($L_\mathrm{150MHz}>L_\textrm{pl}$) and SF-dominated ($L_\mathrm{150MHz}<L_\textrm{eq}$) - based on the dominant component in quasar radio emission instead of radio loudness. Our results suggest that previous disagreements in the inferred dependence of radio loudness on BH mass originate from different definitions of RL/RQ quasars applied in their analysis, leading to contamination of the RL sample by SF-dominated sources, which significantly weaken any observed dependence. Our new definition, motivated by the actual physical model, can reconcile this disagreement. Through our analysis, the origin of the radio excess correlated with the BH mass can be attributed to just the most massive quasars (with top 20\% BH mass) and affect only the most radio-loud (with top 5\% radio loudness) quasars dominated by AGN jets, at a given redshift and bolometric luminosity. We further prove that the differences in this particular quasar population are not driven by bias in BH mass measurements, and these quasars also show differences in their Eddington ratios and their distribution in the \ion{C}{iv} space. \par

We present potential scenarios that could explain these observed differences, relating to black hole accretion mode changes, BH spin, quasar environment, or observational biases in the $M_\mathrm{BH}$ measurements driven by outflows. Distinguishing between these scenarios is not yet possible, but remains of huge interest for future studies.

\section*{Acknowledgements}

BY would like to thank for the support from the University of Edinburgh and Leiden Observatory through the Edinburgh-Leiden joint studentship. KJD acknowledges funding from the European Union's Horizon 2020 research and innovation programme under the Marie Sk\l{}odowska-Curie grant agreement No. 892117 (HIZRAD) and support from the STFC through an Ernest Rutherford Fellowship (grant number ST/W003120/1). LKM is grateful for support from the Medical Research Council [MR/T042842/1]. Several authors are grateful for support from the UK Science and Technology Facilities Council (STFC) via grants ST/V000594/1 (PNB), ST/Y000951/1 (PNB), ST/V000624/1 (MIA), ST/T506047/1 (JP), ST/V506643/1 (JP), ST/X508408/1 (SS), ST/V000624/1 (DJBS), and ST/Y001028/1 (DJBS). BY gratefully acknowledges Yuming Fu for the useful discussions on PyQSOFIT.\par

For the purpose of open access, the author has applied a Creative Commons Attribution (CC BY) licence to any Author Accepted Manuscript version arising from this submission. \par

LOFAR data products were provided by the LOFAR Surveys Key Science project (LSKSP; \url{https://lofar-surveys.org/}) and were derived from observations with the International LOFAR Telescope (ILT). LOFAR \citep[][]{van_haarlem_lofar_2013}{}{} is the Low Frequency Array designed and constructed by ASTRON. It has observing, data processing, and data storage facilities in several countries, which are owned by various parties (each with their own funding sources), and which are collectively operated by the ILT foundation under a joint scientific policy. The efforts of the LSKSP have benefited from funding from the European Research Council, NOVA, NWO, CNRS-INSU, the SURF Co-operative, the UK Science and Technology Funding Council and the J\"ulich Supercomputing Centre. This research made use of the University of Hertfordshire high-performance computing facility and the LOFAR-UK computing facility located at the University of Hertfordshire and supported by STFC [ST/P000096/1].\par

Funding for the Sloan Digital Sky Survey IV has been provided by the Alfred P. Sloan Foundation, the U.S. Department of Energy Office of Science, and the Participating Institutions. \par

SDSS-IV acknowledges support and resources from the Center for High Performance Computing at the University of Utah. The SDSS website is \url{www.sdss4.org}. \par

SDSS-IV is managed by the Astrophysical Research Consortium for the Participating Institutions of the SDSS Collaboration including the Brazilian Participation Group, the Carnegie Institution for Science, Carnegie Mellon University, Center for Astrophysics | Harvard \& Smithsonian, the Chilean Participation Group, the French Participation Group, Instituto de Astrof\'isica de Canarias, The Johns Hopkins University, Kavli Institute for the Physics and Mathematics of the Universe (IPMU) / University of Tokyo, the Korean Participation Group, Lawrence Berkeley National Laboratory, Leibniz Institut f\"ur Astrophysik Potsdam (AIP),  Max-Planck-Institut f\"ur Astronomie (MPIA Heidelberg), Max-Planck-Institut f\"ur Astrophysik (MPA Garching), Max-Planck-Institut f\"ur Extraterrestrische Physik (MPE), National Astronomical Observatories of China, New Mexico State University, New York University, University of Notre Dame, Observat\'ario Nacional / MCTI, The Ohio State University, Pennsylvania State University, Shanghai Astronomical Observatory, United Kingdom Participation Group, Universidad Nacional Aut\'onoma de M\'exico, University of Arizona, University of Colorado Boulder, University of Oxford, University of Portsmouth, University of Utah, University of Virginia, University of Washington, University of Wisconsin, Vanderbilt University, and Yale University.

\section*{Data Availability}

The datasets used in this paper were derived from sources in the public domain: the LOFAR Two-Metre Sky Surveys (\url{www.lofar-surveys.org}) and the Sloan Digital Sky Survey (\url{www.sdss.org}). Solutions to $L_\mathrm{eq}$ and $\sigma_\mu$ in $\mathcal{M}_i-z$ grid cells not included in this work can be provided upon reasonable request to the lead author.



\bibliographystyle{mnras}
\bibliography{references} 



\clearpage
\appendix

\section{Characteristic luminosities of SF- and AGN-dominated quasars}
\label{sec:appendix_result}

In Table~\ref{tab:lum_values}, we present the values of the characteristic 150 MHz radio luminosities $L_\textrm{eq}$ and $L_\textrm{pl}$ (and their corresponding flux densities $f_\mathrm{eq}$ and $f_\mathrm{pl}$, as evaluated from the midpoint of each redshift bin) defined in Section~\ref{sec:analysis}, for the $\mathcal{M}_i-z$ grid cells included in this work. These values can be applied directly to classify quasars from other sources as long as their rest-frame \emph{i}-band magnitudes and redshifts lie within the grid cells and 150 MHz (or equivalent) radio flux densities measurements are available. Following the instructions in Section~\ref{sec:analysis}, readers can also calculate $L_\textrm{eq}$ and $L_\textrm{pl}$ for the rest of the grid cells in \citetalias{yue_novel_2024} (while not included in this work) from the model best fits presented in Table A1 of \citetalias{yue_novel_2024}. \par

\begin{table}
    \centering
\caption{The characteristic 150 MHz radio luminosity $L_\textrm{eq}$ (upper luminosity threshold for SF-dominated quasars) and $L_\textrm{pl}$ (lower luminosity threshold for AGN-dominated quasars) in the $\mathcal{M}_i-z$ grid cells covered by this work. The $\mathcal{M}_i$ and $z$ values in this table represent the central point for each grid cell, while each grid cell spans 1 mag in rest frame \emph{i}-band magnitude and 0.4 in redshift. The radio flux densities $f_\mathrm{eq}$ and $f_\mathrm{pl}$ are calculated from the midpoint redshift of each $\mathcal{M}_i-z$ grid cell.}\label{tab:lum_values}
    \begin{tabular}{cccccc}
        \hline
        $M_i$  & $z$ & $\log L_\mathrm{eq}$  &$f_\mathrm{eq}$& $\log L_\mathrm{pl}$  &$f_\mathrm{pl}$\\
         $(z=0)$& & $[\mathrm{W~Hz}^{-1}]$  &$[\mathrm{mJy}]$& $[\mathrm{W~Hz}^{-1}]$  &$[\mathrm{mJy}]$\\
        \hline
        -21.5 & 0.6 & 23.86 & 0.541 & 24.19 & 1.152 \\
        -22.5 & 0.6 & 24.23 & 1.261 & 24.66 & 3.406 \\
        -23.5 & 0.6 & 24.41 & 1.890 & 24.98 & 7.104 \\
        -22.5 & 1.0 & 24.50 & 0.704 & 24.81 & 1.455 \\
        -23.5 & 1.0 & 24.65 & 0.998 & 25.07 & 2.610 \\
        -24.5 & 1.0 & 24.64 & 0.973 & 25.18 & 3.397 \\
        -22.5 & 1.4 & 24.66 & 0.463 & 24.95 & 0.898 \\
        -23.5 & 1.4 & 24.86 & 0.734 & 25.22 & 1.700 \\
        -24.5 & 1.4 & 24.94 & 0.891 & 25.38 & 2.400 \\
        -25.5 & 1.4 & 25.02 & 1.051 & 25.55 & 3.639 \\
        -23.5 & 1.8 & 24.97 & 0.525 & 25.23 & 1.104 \\
        -24.5 & 1.8 & 25.16 & 0.815 & 25.55 & 2.017 \\
        -25.5 & 1.8 & 25.23 & 0.941 & 25.70 & 2.825 \\
    \end{tabular}
\end{table}

Since the definitions of $L_\textrm{eq}$ and $L_\textrm{pl}$ are independent of the quasar selection or the parameter space specific to this work, our classification method can be extended to other parameter spaces (defined with $\mathcal{M}_i-z$ or other physical properties) as long as a best fit of the two-component model in \citetalias{yue_novel_2024} ($\{\Psi,\sigma_\Psi,\gamma,f\}$) can be obtained from a certain distribution of quasar radio flux densities.

\section{Additional plot for black hole mass distributions}
\label{sec:appendix_ecdf}

\begin{figure*}    
    \includegraphics[width=\textwidth]{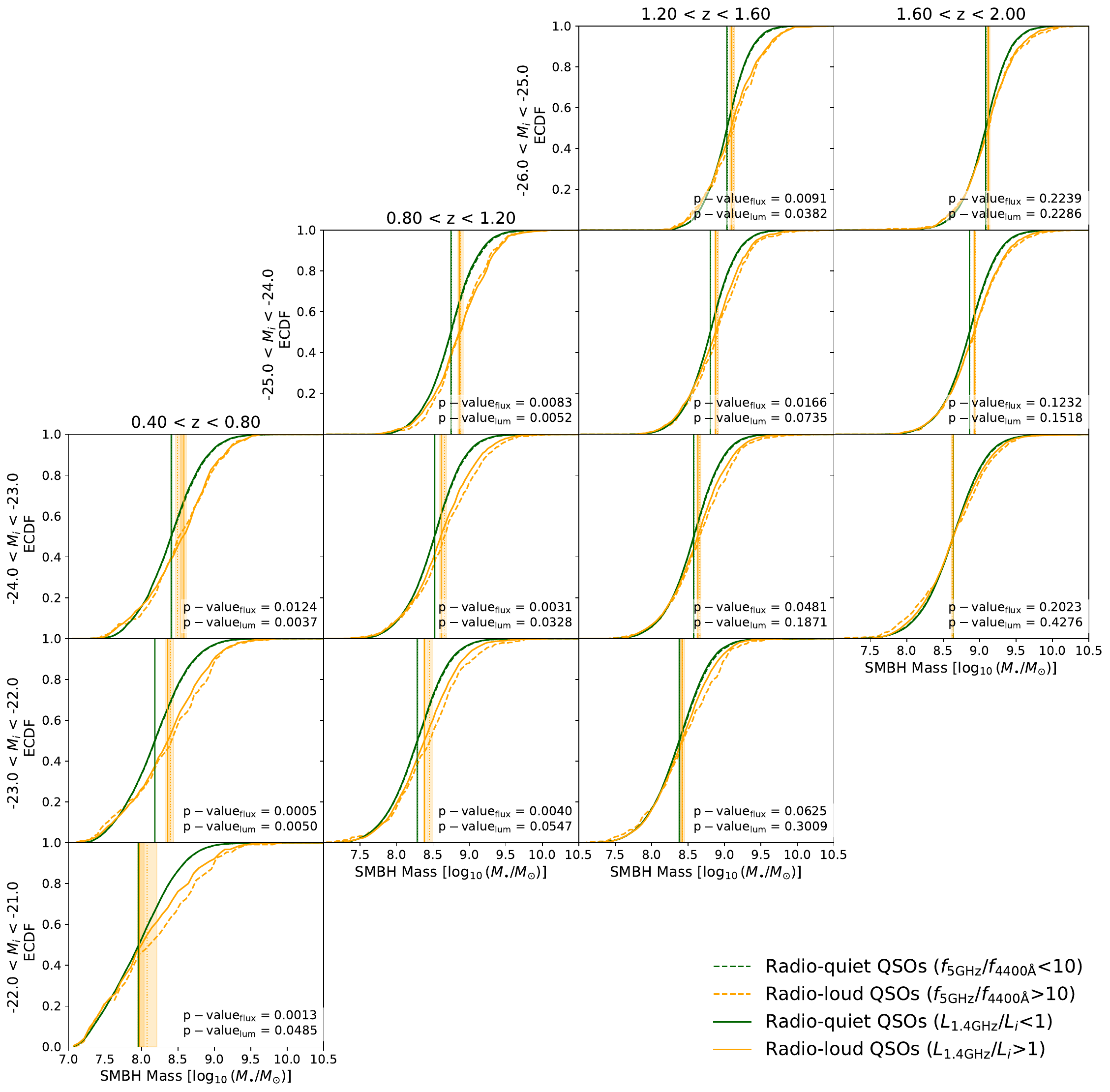}
    \caption{Cumulative mass distribution within each $\mathcal{M}_i-z$ grid cell, separated by the radio-loudness of quasars. The grid cells are arranged by their average \emph{i} band luminosity and redshift, following the pattern in Figure~\ref{fig:ecdf_mass_with_fit}. For the distribution drawn in solid lines, the RL/RQ classification is based on the values of $R_\mathrm{lum}=\log_{10}\left(L_\textrm{1.4GHz}/L_i\right)$: quasars with $R_\mathrm{lum}<1$ are classified as RQ quasars, while quasars with $R_\mathrm{lum}>1$ are classified as RL. The distribution in dashed lines reflects the RL and RQ quasar population defined with $R_\mathrm{flux}=f_\textrm{5GHz}/f_\textrm{4400\AA}$, where the RQ quasars have $R_\mathrm{flux}<10$ and the RL quasars have $R_\mathrm{flux}<10$. Within each grid cell, we include the p-values calculated from K-S tests on the BH mass CDFs of RL and RQ quasars under different definitions.}
    \label{fig:ecdf_mass}
\end{figure*}

Figure~\ref{fig:ecdf_mass} shows the cumulative BH mass distribution across the $\mathcal{M}_i-z$ parameter space for RL and RQ quasar populations defined in the previous literature, using $R_\mathrm{flux}=f_\textrm{5GHz}/f_\textrm{4400\AA}$ \citep[e.g.][]{mclure_relationship_2004} and $R_\mathrm{lum}=\log_{10}\left(L_\textrm{1.4GHz}/L_i\right)$ \citep[e.g.][]{balokovic_disclosing_2012,arnaudova_exploring_2024}. Based on both criteria, the BH mass distribution differs between the RL and RQ samples, where the RL quasars tend to be more massive (p-value<0.05) in some grids; however, there is no unified trend in such a difference, suggesting that radio loudness alone is not enough to characterise the relationship between quasar radio emission and $M_\mathrm{BH}$. According to the criterion used by \citet{arnaudova_exploring_2024} where quasars with $R_\mathrm{lum}<1$ are classified as RQ quasars and quasars with $R_\mathrm{lum}>1$ are classified as RL (dotted lines), there is little difference (p-value>0.05) between the BH mass distributions of RL and RQ quasars in many of the more-populated-lower-luminosity grid cells, which leads to the disagreement between their conclusion and the result in \citet{mclure_relationship_2004}. For the less populated grid cells in the higher luminosity regime, both criteria agree with each other due to the different host galaxy SFR and AGN jet normalisation in those parameter space.

\label{lastpage}
\end{document}